\documentclass[preprint,3p,twocolumn]{elsarticle}
\usepackage{geometry}
\usepackage{fancyhdr}
\biboptions{numbers,sort&compress}

\usepackage[usenames,dvipsnames]{xcolor}
\usepackage{graphicx}
\usepackage{amsmath}
\usepackage{amssymb}
\usepackage{verbatim}
\usepackage{wasysym}
\usepackage{multirow}
\usepackage{hyperref}
\usepackage{cancel} 
\usepackage{soul}

\biboptions{comma,square,sort&compress}
\setstcolor{red}

\begin{document}
\begin{frontmatter}
 \title{Grain growth competition and formation of grain boundaries during solidification of hcp alloys}
 \author{A.K. Boukellal$^{1,\dag,*}$}
 \author{M. Sarebanzadeh$^{1,2}$}
 \author{A. Orozco-Caballero$^{3}$}
 \author{F. Sket$^{1}$}
 \author{J. LLorca$^{1,2}$}
 \author{D. Tourret$^{1,*}$}
 \address{$^1$ IMDEA Materials Institute, Getafe, 28906 Madrid, Spain.}
 \address{$^2$ Department of Materials Science, Polytechnic University of Madrid/Universidad Polit\'ecnica de Madrid, 28040 Madrid, Spain.}
\address{$^3$ Department of Mechanical Engineering, Chemistry and Industrial Design, Polytechnic University of Madrid/Universidad Polit\'ecnica de Madrid, 28012 Madrid, Spain.}
 \cortext[cor1]{\raggedright
 Corresponding authors; Email addresses: ahmed.boukellal@univ-lorraine.fr, damien.tourret@imdea.org. \\
 $^\dag$ Present address: Universit\'e de Lorraine, CNRS -- IJL, Nancy, France.
 }

\begin{abstract}

Grain growth competition during directional solidification of a polycrystal with hexagonal (hcp) symmetry (Mg-1wt$\%$Gd alloy) is studied by phase-field modeling, exploring the effect of the temperature gradient $G$ on the resulting grain boundary (GB) orientation selection.
Results show that selection mechanisms and scaling laws derived for cubic (fcc, bcc) crystals also apply to hcp materials (within their basal plane), provided a re-estimation of fitting parameters and re-scaling to account for the sixfold symmetry.
While grain growth competition remains stochastic with rare events of unexpected elimination or side-branching along the developing GBs, we also confirm an overall transition from a geometrical limit to a favorably oriented grain limit  behavior with an increase of thermal gradient within the dendritic regime, and the progressive alignment of dendrites and GBs toward the temperature gradient direction with an increase of $G$ during the dendritic-to-cellular morphological transition.
Comparisons with original thin-sample directional solidification experiments show a qualitative agreement with PF results, yet with notable discrepancies, which nonetheless can be explained based on the stochastic variability of selected GB orientations, and the statistically limited experimental sample size.
Overall, our results extend the understanding of GB formation and grain growth competition during solidification of hcp materials, and the effect of thermal conditions, nonetheless concluding on the challenges of extending the current studies to three dimensions, and the need for much broader (statistically significant) data sets of GB orientation selected under well-identified solidification conditions.

\end{abstract}
\end{frontmatter}
\thispagestyle{fancy} 
\section{Introduction}

Metals and alloys represent a crucial class of structural materials in the manufacturing sector. 
Their crystalline microstructures and mechanical behavior are relatively well-identified and understood within a single crystal \cite{kurz84, dantzig16,kurz2019progress,kurz2021progress}. 
Yet, while most technological metallic materials are polycrystalline, a fundamental understanding remains lacking upon key mechanisms of the formation of heterogeneous polycrystalline microstructures and upon their resulting mechanical behavior, in particular for materials with complex hexagonal (hcp) crystal structures. 
Among these, Magnesium-based alloys represent tremendous opportunities for weight reduction in automotive applications \cite{mordike2001magnesium, taub2007evolution, luo2013magnesium}, as well as a strong potential in aeronautics and, most recently, in biodegradable implants \cite{hort2010magnesium, chen2014recent}. Mg-rare earth (Mg-RE) alloys, in particular, have recently emerged with improved mechanical properties, especially hardness, yield strength, tensile and creep resistance \cite{hort2010magnesium,yang09,peng12,xu18}. 
The mechanical properties can be improved by heat treatment processes, but also by controlling the microstructure during the solidification process, especially when macrosegregation-related phenomena are involved \cite{kurz84,dantzig16}. 

Microstructural characteristics, such as dendritic spacings and polycrystalline grain texture, determine the properties, performance, and lifetime of metal cast parts. 
Most often, solidification processing is the first processing step accompanied by microstructure formation. 
Even though our understanding of crystal growth and solidification as a whole has evolved significantly over the past decades \cite{boettinger2000solidification, asta2009solidification, kurz2019progress, kurz2021progress}, microstructure selection mechanisms are still not completely understood. 
For instance, a comprehensive picture of dendritic grain growth competition, i.e., of the selection of the shape and orientation of grain boundaries (GBs), is still lacking, in particular for hcp materials.

The overall objective of this work is to establish a comprehensive understanding of microstructure selection and grain growth competition during solidification of hcp alloys. 
To do so, we perform two-dimensional (2D) phase-field (PF) simulations of directional solidification of a Mg-1wt$\%$Gd alloy, using a thoroughly-validated quantitative PF model \cite{karma01,echebarria04} with an hcp anisotropy vector formulation \cite{boukellal21}, accelerated via parallelization using graphics processing units (GPUs).
We perform a full mapping of the bi-crystalline grain orientation, exploring a wide range of temperature gradient $G$, and compare our results to current theories for the selection of primary dendrite arm spacing (PDAS), dendritic growth orientation, and GB orientation selection, so far established within the context of cubic (fcc, bcc) crystal structures.
We also compare our simulation results with original experimental measurements in thin samples directionally solidified under well-identified temperature gradient $G$ and growth velocity $V$.

\section{State-of-the-art theories}
\label{Sec:theory}

Before presenting our methods (Sec.~\ref{Sec:method}) and discussing our results (Sec.~\ref{Sec:result}), first we review the state-of-the-art, in particular theories for the selection of PDAS (Sec.~\ref{sec:theory:pdas}), growth orientation (Sec.~\ref{sec:theory:dgp}), and GB orientation selection during directional solidification of alloys (Sec.~\ref{sec:theory:theta}).

\subsection{Selection of primary dendrite arm spacing}
\label{sec:theory:pdas}

For a given set of processing parameters $(G,V)$, a wide range of dendritic array spacings may be selected \cite{Kurz81,somboonsuk85,trivedi85,lu92,warren93,han94,hunt96}. 
In the case of dendrites growing in a direction tilted by an angle $\alpha$ with respect to the temperature gradient $G$, we denote $\Lambda_0$ the primary dendrite arm spacing (PDAS) measured in the normal direction to the temperature gradient, while $\Lambda_\alpha$ stands for the PDAS measured normal to the dendrite growth direction (later illustrated in Figure~\ref{Fig:Imp}).
Classical phenomenological models typically predict $\Lambda_0$ in an array of dendrites aligned with $G$, as a function of the processing parameters \cite{trivedi94}. 
Gandin et al. \cite{gandin96} proposed an orientation-dependent PDAS model         

\begin{equation}
\Lambda_\alpha(G,\alpha)=LG^{-c}\{1+d[\cos(\alpha)^{-e}-1]\} 
\label{Eq:Gand}
\end{equation}
that accounts for the growth direction angle $\alpha$ with respect to $G$. 
The prefactor $L$ depends on the alloy properties (e.g. phase diagram) and growth velocity $V$ (here a constant) while $c$, $d$ and $e$ are used as fitting parameters.

The resulting orientation-dependence of the PDAS was validated against experimental measurements \cite{gandin96} and phase-field simulations \cite{tourretkarma15} for thin-sample and regular arrays \cite{song2023cell} directional solidification of cubic-symmetry systems (e.g transparent succinonitrile compounds).
Yet, to the best of our knowledge, it was not compared to either experimental or computational results for systems with a hcp crystal structure.

\subsection{Preferred crystal growth direction}
\label{sec:theory:dgp}

At low velocity, just above the onset of morphological destabilization of a planar solid-liquid interface, cellular patterns exhibit a mostly isotropic shape and grow along the main temperature gradient direction.
At higher velocities, dendrite orientation transitions have been reported in various systems as a function of alloy composition \cite{haxhimali2006orientation, dantzig2013dendritic, becker2019dendrite, wang2020controlling} or cooling rates \cite{bedel2015characterization}.
Still, most crystalline solids have been reported to grow along well-defined crystallographic directions. 
For cubic systems (fcc, bcc), the preferred growth direction is usually $\langle100\rangle$, while for hexagonal symmetry (hcp) systems, both $\langle10\bar10\rangle$ (most common in Zn \cite{strutzenberger1998solidification, semoroz2002orientation}) and $\langle11\bar20\rangle$ (most common in Mg \cite{pettersen1989crystallography, pettersen1990dendritic}) preferred growth directions have been reported.

Between cellular and dendritic regimes, the preferred growth direction $\alpha$ shows a continuous transition from the orientation of the thermal gradient to the orientation selected by the crystal structure. 
This transition has been formulated as \cite{akamatsu97,dgp08}

\begin{equation}
\frac{\alpha}{\alpha_0}=1-\frac{1}{1+f\,\rm{Pe}^g}
\label{Eq:DGP}
\end{equation}     
expressing the ratio between the growth direction $\alpha$ and the crystal lattice (e.g. $\langle100\rangle$) orientation $\alpha_0$ as a function of the spacing Péclet number, ${\rm Pe} =\Lambda_0 V/D$.
The latter stands as a measure of the interactions between the neighboring dendrites via the solutal diffusion field. 
Here, $\Lambda_0$ is the PDAS measured perpendicularly to $G$, and $f$ and $g$ are two fitting parameters. 
At high Pe numbers, i.e., high PDAS and/or low $G$, $\alpha \approx \alpha_0$ (i.e. growth along crystallographic directions in the dendritic regime), while $\alpha \approx 0^\circ$ at low Pe numbers, i.e.,  low PDAS and/or high $G$ (i.e. growth aligned with $G$ in the cellular regime).

This progressive transition has been derived from observations in thin-sample directional solidification experiments \cite{akamatsu97,dgp08} and was later validated computationally using quantitative phase-field simulations \cite{tourretkarma15, song2018propagative, mota2021effect, song2023cell}.
The latter have shown that the parameter $f$, initially proposed to be a function of $\alpha_0$ \cite{akamatsu97,dgp08}, could be considered a constant for a given material, and that Eq.~\eqref{Eq:DGP} remains valid for three-dimensional systems with different spatial array ordering (see Supplementary Information of \cite{song2023cell}).
However, these studies all focused on cubic-symmetry crystals and, to the best of our knowledge, this growth orientation selection during the cellular to dendritic transition has not been reported for hcp crystals so far.

\subsection{Selection of GB orientations during solidification}
\label{sec:theory:theta}

\begin{figure}[t!]
\includegraphics[width=3in]{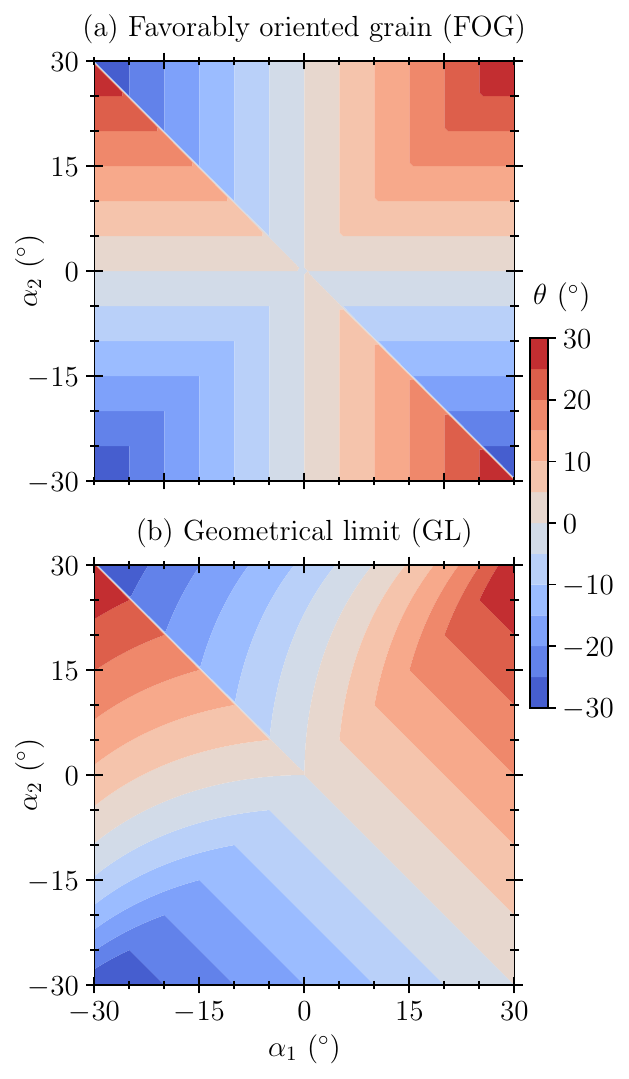}
\caption{
Grain boundary orientation angle $\theta$ shown as a function of the misorientation of grain 1, $\alpha_1$, and grain 2, $\alpha_2$ (see Fig.~\ref{Fig:Imp}) using (a) the Favorably Oriented Grain (FOG) criterion (Eq.~\eqref{Eq:FOG}) and (b) the Geometrical Limit (GL) (Eq.~\eqref{Eq:GL}).   
For simplicity, both maps consider a well-developed dendritic regime with $\alpha\approx\alpha_0$ (see Section~\ref{sec:theory:dgp}).
}
\label{Fig:FOG-GL_abs}
\end{figure}

\begin{figure}[!t]
\includegraphics[width=3in]{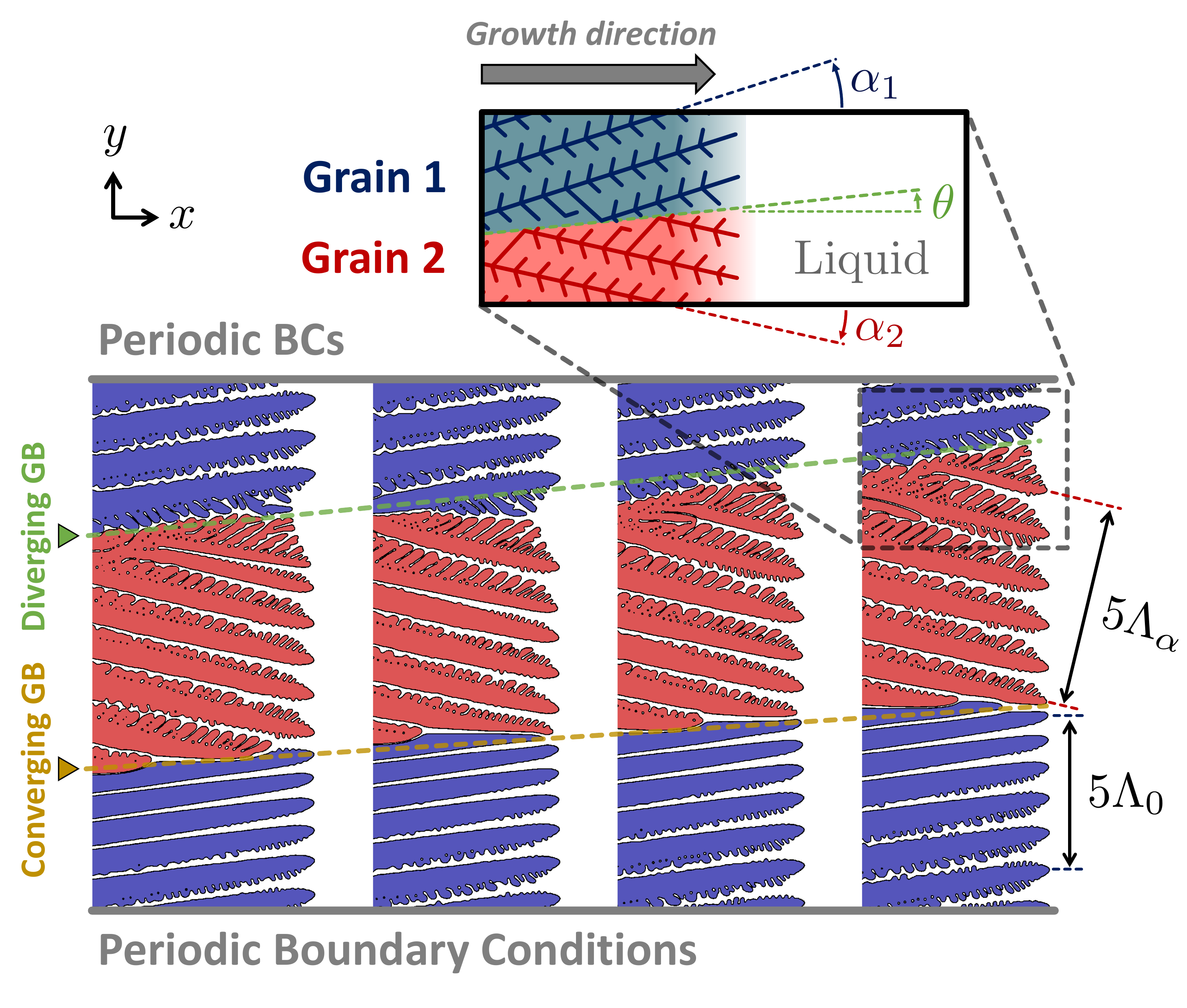}
\caption{
Configuration of the PF simulations showing two, red and blue, grains growing from left to right (direction of the temperature gradient, $G$).
Periodic boundary conditions applied normally to the growth direction onto to a bicrystalline domain (bottom) allow tracking two GBs, one converging GB and one diverging GB. 
For each GB (top schematic), angles $\alpha_1$ and $\alpha_2$ are defined as the orientation of the upper and lower grains, respectively, and $\theta$ is the orientation of the GB.
All angles are measured counterclockwise with respect to the thermal gradient $G$ main direction (here aligned with the $x$ axis), i.e. here illustrated for $\alpha_1>0$, $\alpha_2<0$ and $\theta>0$. 
In a grain, PDAS $\Lambda_0$ and $\Lambda_\alpha$ are measured perpendicular to $G$ and to the growth direction, respectively.
}
\label{Fig:Imp}
\end{figure}

The early classical theory of grain growth competition by Walton and Chalmers \cite{chalmers59} states that the grain best aligned with the thermal gradient direction, which grows at a lower undercooling compared to grains with higher misorientation with respect to $G$, cannot be eliminated by these higher-misorientation neighbors. 
A common extrapolation of this argument leads to the consideration that the GB orientation equals the growth orientation of the best-oriented grain, i.e. the minimum $|\alpha|$ among the two grains.
This is known as the favorably oriented grain (FOG) criterion.
The resulting GB orientation $\theta$, illustrated by the $\theta(\alpha_1,\alpha_2)$ map in Fig.~\ref{Fig:FOG-GL_abs}a, is thus given by 

\begin{equation}
\theta_{\rm FOG}(\alpha_1,\alpha_2)=\begin{cases}
    \alpha_1 & \quad \text{if } |\alpha_1|\leq|\alpha_2| \\
    \alpha_2 & \quad \text{if } |\alpha_2|\leq|\alpha_1|
  \end{cases}
\label{Eq:FOG}  
\end{equation}
where $\alpha_1$ and $\alpha_2$ are the growth directions of the competing grains labelled 1 and 2.  
As illustrated in Figure~\ref{Fig:Imp}, to be consistent with previous publications \cite{tourretkarma15,tourret17,pineau18, dorari22} we consider that grain 1 is above the GB while grain 2 is below the GB if the growth direction goes from left to right.
Early experimental observations supported qualitatively this simple description by Eq.~\eqref{Eq:FOG}, e.g. exhibiting progressive elimination of grains with the highest misorientation with $G$ \cite{esaka85,esaka86}.
However, phenomena contradicting this simple theory, such as the unusual overgrowth of well-oriented dendrites at converging GB, were observed experimentally \cite{wagner04,zhou08} and later reproduced and explained with phase-field simulations \cite{li12,takaki16,tourret17}. 

Pineau et al. \cite{pineau18} proposed another asymptotic growth limit, termed the geometrical limit (GL), which can be interpreted as considering continuous side-branching, i.e. infinitely small secondary arm spacing, such that side-branches instantaneously occupy the space between two grains.
The resulting GB orientation, illustrated by the $\theta(\alpha_1,\alpha_2)$ map in Fig.~\ref{Fig:FOG-GL_abs}b, is thus expressed as \cite{pineau18}

\begin{equation}
\theta_{\rm GL}(\alpha_1,\alpha_2)=
\begin{cases}
& \tan^{-1}\Big(\frac{\cos \alpha_1 +\sin \alpha_1 - \sec \alpha_2}{\cos \alpha_1 -\sin \alpha_1} \Big)  \\
& \quad\quad \text{if } (\alpha_1 +\alpha_2) \geq 0 ~\text{ and }~ (\alpha_1<\alpha_2) \\
& \tan^{-1}\Big(\frac{\cos \alpha_2 +\sin \alpha_2 - \sec \alpha_1}{\cos \alpha_2 +\sin \alpha_2} \Big)  \\
&\quad\quad \text{if } (\alpha_1 +\alpha_2) < 0 ~\text{ and }~ (\alpha_1<\alpha_2) \\
& \frac{\alpha_1+\alpha_2}{2} 
\quad\quad \text{if } (\alpha_1 \geq \alpha_2)
\end{cases}
\label{Eq:GL}  
\end{equation}
where the first and second cases refer to converging GBs ($\alpha_1<\alpha_2$) on either side of the $\{\alpha_1=-\alpha_2\}$ symmetric configuration, while the third case refers mostly to diverging GBs ($\alpha_1>\alpha_2$). 

Selection maps of $\theta(\alpha_1,\alpha_2)$ are expected to be periodic in both $\alpha_1$ and $\alpha_2$, with a period $90^\circ$ for fourfold (fcc,bcc) or $60^\circ$ for sixfold (hcp in the basal plane) symmetry.
This periodicity appears in both FOG (Fig.~\ref{Fig:FOG-GL_abs}a) and GL (Fig.~\ref{Fig:FOG-GL_abs}b) maps.
Moreover, both maps exhibit a discontinuity along the $\{\alpha_1=-\alpha_2\}$ diagonal.
Indeed, from symmetry considerations, once averaged over a statistically significant number of GBs, it is expected that $\theta(\alpha_1=-\alpha_2)=0^\circ$.
The same discontinuity, though less visible in Fig.~\ref{Fig:FOG-GL_abs}, occurs at the $\{\alpha_1=\alpha_2=\pm30^\circ\}$ corners. 
This latter discontinuity is even more pronounced when differences between $\alpha$ and $\alpha_0$ are considered (cf. Sec.~\ref{sec:theory:dgp}).
While FOG and GL theories both consider dendritic microstructures, actual microstructures usually transition toward a degenerate (i.e., seaweed-like) patterns as they approach $\alpha_0\to\pm30^\circ$ ($\pm45^\circ$ in fourfold symmetry systems).
The discontinuity at the four corners of the $\theta(\alpha_1,\alpha_2)$ plot is thus usually replaced by a transition region within a $(\alpha_1,\alpha_2)$ range directly linked to the dendritic-to-degenerate transition region \cite{tourretkarma15,tourret17}.

GB orientation data across a broad range of crystalline orientations $(\alpha_1,\alpha_2)$ is tedious to gather experimentally and hence scarce, such that the selection of $\theta(\alpha_1,\alpha_2)$ has been approached primarily by phase-field simulations. 
The most extensive mapping of $(\alpha_1,\alpha_2)$ space in bi-crystalline samples to date \cite{tourret17} has shown that the GB orientation selection deviates from the classical FOG description, Eq.~\eqref{Eq:FOG}, in the following ways: (i) $|\theta|$ remains low for diverging GBs with $\alpha_1\times\alpha_2<0$ (hence closer to the GL in the bottom right quadrant of Fig.~\ref{Fig:FOG-GL_abs}), (ii) the PF selection map exhibits a narrow transition region along the symmetric $\{\alpha_1=-\alpha_2\}$ configuration, and (iii) a transition to degenerate seaweed-like patterns may occur toward the $\{\alpha_i=45^\circ\}$ ($i=1,2$) symmetric configuration (in fourfold cubic symmetry crystals, hence with singularity at $45^\circ$).

Several studies have highlighted the important role of the temperature gradient $G$ \cite{tourretkarma15,takaki16,dorari22}, in particular close to the cellular-to-dendritic transition regime \cite{tourretkarma15,takaki16}, and also suggestted a transition from mostly-FOG to mostly-GL behavior with a decrease in $G$ in the dendritic regime \cite{dorari22}.
Recent studies have also highlighted the morphological complexity of three-dimensional (3D) GBs \cite{takaki18,song2023cell}, as observed in directionally solidified samples \cite{hu2018competitive}.
However, given the added dimensionality of the 3D GB bi-crystal problem, the computational cost of 3D simulations still stands as bottleneck to perform a comprehensive mapping of grain orientations and processing conditions $(G,V)$.
Moreover, except a handful of phase-field simulations by Eiken \cite{eiken10,eiken2010phase} (to our knowledge, the earliest reports of PF-simulated grain growth competition), the problem of bi-crystal grain growth competition in hcp crystals remains for the most part unexplored.

\section{Methods}
\label{Sec:method}

Here, for the sake of reproducibility, we present details on the PF model (Sec.~\ref{sec:meth:pf}) and simulations (Sec.~\ref{sec:meth:simu}), as well as solidification experiments (Sec.~\ref{sec:meth:exp}). 
With the exception of the hcp six-fold symmetry of the interface excess free energy (implemented using an auxiliary anisotropy vector \cite{boukellal21}), the model and simulation setup are similar to what was already employed in previous studies \cite{tourretkarma15,tourret17}.

\subsection{Phase-field model}
\label{sec:meth:pf}

We use the classical quantitative thin-interface phase-field model for directional solidification of a binary alloy with anti-trapping current \cite{echebarria04}, a preconditioned phase field $\psi=\sqrt{2}\tanh^{-1}(\varphi)$ \cite{Glasner01,tourretkarma15}, and a formulation and implementation including the anisotropy vector $\vec A$ to facilitate the use of arbitrary interface energy anisotropy functions \cite{boukellal21}.
The resulting evolution equations for the preconditioned phase field $\psi$ and dimensionless solute concentration (supersaturation) field $U$ are 

\begin{equation}
\begin{split}
\Bigg[1- & (1-k) \frac{x-\tilde{V}t}{\tilde{l_T}} \Bigg] \frac{\partial \psi}{\partial t} = \\
& \sqrt{2}\Bigg[\varphi -\lambda\Big(1-\varphi^2\Big)\Big(U+\frac{x-\tilde{V}t}{\tilde{l}_T}\Big)\Bigg] \\
& + a_s^2\Big[ \nabla^2\psi -\sqrt{2}\varphi\vert\vec{\nabla}\psi \vert^2\Big]\\
& + 2a_s\vec{\nabla}a_s\vec{\nabla}\psi 
+\frac{\sqrt{2}}{1-\varphi ^2}\vec{\nabla}\cdot\vec{A}
\end{split}
\label{pf1}
\end{equation}

\noindent
and 

\begin{align}
\begin{split}
[1 +&  k -(1-k) \varphi]\frac{\partial U}{\partial t} =\tilde{D}\vec{\nabla}\Big[ (1-\varphi)\vec{\nabla}U\Big]\\ 
& +\Big[ 1+(1-k)U \Big]\frac{(1-\varphi ^2)}{\sqrt{2}}\frac{\partial \psi}{\partial t}\\
& +\vec{\nabla}\left\{\Big[1+(1-k)U\Big]\frac{(1-\varphi^2)}{2}\frac{\partial\psi}{\partial t}\frac{\vec{\nabla }\psi}{\vert \vec{\nabla}\psi\vert}\right\}
\end{split}
\label{u1}
\end{align}

\noindent
where 

\begin{equation}
U=\frac{1}{1-k}\left[\frac{c/c_l^0}{(1-\varphi)/2+k(1+\varphi)/2}-1\right]
\end{equation}

\noindent
with $c_l^0=c_\infty/k$ the solute concentration on the liquid side of a flat interface at the solidus temperature $T_0$ for an alloy of nominal concentration $c_\infty$.
This model considers a linearized phase diagram (or dilute alloy limit) with constant liquidus slope $m$ and interface solute partition coefficient $k$, a solute diffusion coefficient $D$ in the liquid phase and negligible diffusion in the solid phase.
The temperature field follows the common ``frozen temperature approximation'' with constant temperature gradient $G$ and pulling (i.e. isotherms) velocity $V$ in the $x$ direction, resulting in the one-dimensional temperature field $T(x)=T_0+G(x-Vt)$.
The interface thickness $W$ and the phase field equation relaxation time $\tau_0$ are taken as space and time units respectively \cite{karma01,echebarria04}. 
Thus, the dimensionless form of 
the diffusion coefficient $\tilde{D}$, the growth velocity $\tilde{V}$ and the thermal length $\tilde{l}_T$
are given respectively by 

\begin{equation}
\tilde{D}=\frac{D\tau_0}{W^2}=a_1a_2\frac{W}{d_0},
\end{equation}
\begin{equation}
\tilde{V}=\frac{V\tau_0}{W}=a_1a_2\frac{Vd_0}{D}\Big(\frac{W}{d_0}\Big),
\end{equation}

\noindent
and

\begin{equation}
\tilde{l}_T=\frac{l_T}{W}=\frac{l_T}{d_0}\frac{1}{W/d_0},
\end{equation}

\noindent
with $d_0=\Gamma/[mc_\infty(1-1/k)]$ the chemical capillary length at $T_0$,
$\Gamma$ the interface Gibbs-Thomson coefficient, 
$l_T=[mc_\infty(1-1/k)]/G$ the thermal length, 
$a_1=5\sqrt{2}/8$ and $a_2=47/75$ \cite{karma01,echebarria04}.
In the bulk liquid ($\varphi=-1$), Eq.~\eqref{u1} reduces to the solute diffusion equation, whereas at the interface using a coupling coefficient $\lambda =a_1W/d_0$ between the phase field and the supersaturation ensures a vanishing kinetic undercooling \cite{karrap98,echebarria04}. 
In Eq.~\eqref{pf1}, $a_s(\vec n) = \gamma(\vec n) / \gamma_0  $ expresses the anisotropy of the interface excess free energy projected on the basal plane, formulated as \cite{sun06,eiken2010phase}

\begin{equation}
a_s(\vec{n}) = \Big[ 1.008 \,+\,  \varepsilon_{66}\frac{\sqrt{6006}}{64\sqrt{\pi}} (n_x^6-15n_x^4n_y^2+15n_x^2n_y^4-n_y^6))          
\Big]
\label{pf_aniso}
\end{equation} 

\noindent
where $\varepsilon _{66}$ is the strength of the surface energy anisotropy in the basal plane and $(n_x,n_y)$ is the normal vector to the solid-liquid interface calculated as $\vec{n}=-\vec{\nabla}\psi/\lvert\vec{\nabla}\psi\lvert$.
In Eq.~\eqref{pf_aniso}, we consider the two-dimensional sixfold anisotropy in the basal plane with an out-of-plane anisotropy coefficient $\varepsilon_{20}\approx-0.026$ \cite{sun06}.
The auxiliary anisotropy vector $\vec A$ is introduced for convenience and versatility, as it makes it easier to switch between different forms of the interface anisotropy \cite{boukellal21, boukellal2019equilibrium}.
It appears from reformulating the anisotropy (i.e., last) term of the phase-field equation as a divergence 

\begin{align}
\vec\nabla\cdot\vec A = \sum_{i=x,y}{ \partial_i \left[ |\vec\nabla\varphi|^2 a_s \frac{\partial a_s}{\partial \varphi_i} \right] }
\end{align}

\noindent
with 

\begin{align}
A_i 
& =   |\vec\nabla\varphi|^2 a_s \frac{\partial a_s}{\partial \varphi_i} 
\nonumber\\
&= \frac{1-\varphi^2}{\sqrt{2}} |\vec{\nabla}\psi|^2a_s   \frac{\partial a_s}{\partial \psi_\alpha} 
\nonumber\\
& = \frac{1-\varphi^2}{\sqrt{2}}|\vec{\nabla}\psi|a_s\sum_{j=x,y}(n_in_j-\delta_{i,j})\frac{\partial a_s}{\partial n_j}
\label{Eq:Anis_Vec}
\end{align}
with $\delta_{i,j} = 1 $ if $i=j$ and $0$ otherwise.
Details and full expressions of $\vec{A}$, as they appear in the code, are provided in \ref{sec:app:anisovec} (for completeness, also including 3D expressions).

In order to destabilize the initial planar interface and promote side-branching, at each grid point ($i,j$) and each time iteration, we add a random perturbation $\eta \beta_{i,j} \sqrt{\Delta t}$ to the value of $\psi_{i,j}$ at the next time step :

\begin{equation}
\psi_{i,j}(t+\Delta t)=\psi_{i,j}(t)+\Delta t \frac{\partial \psi_{i,j}}{\partial t}+\eta \beta_{i,j} \sqrt{\Delta t}
\label{eq:noise}
\end{equation}
where $\beta_{i,j}$ is a random number picked from a flat distribution in the range $[-0.5,+0.5]$ and $\eta$ is the noise amplitude \cite{karrap99,ghmadh14}.

Equations \eqref{pf1} and \eqref{u1} are solved using finite differences on a regular grid of square elements with a spatial step $\Delta x$. 
They are integrated following an Euler explicit time scheme with a time step $\Delta t = 0.8 \Delta x^2/4\tilde{D} $. 
We use a moving domain following the most advanced dendrite tip in the $x$ direction. 
For performance, Eq. \eqref{Eq:Anis_Vec} is set to zero in the bulk phases, i.e. namely where $(1-\varphi^2) <0.01$.   
The model is written in computer uniﬁed device architecture (CUDA) programming language for massively parallel computing on Graphic Processing Units (GPU).
The time loop is composed of three sequential kernel calls, calculating: (i) the $\vec A$ ﬁeld at the current time step, (ii) the $\psi$ ﬁeld at the next time step, and (iii) the $U$ ﬁeld at the next time step. 
An additional kernel shifting all fields in the $x-$ direction is triggered occasionally when the highest location of the interface in $x$ exceeds a given position $L_{\rm solid}$ in the computational domain.
Time stepping is performed at the end of the time loop by swapping pointer addresses between arrays containing values of $\psi$ and $U$ at the current and next time steps. 

\subsection{Simulations} 
\label{sec:meth:simu}

We study grain growth competition in a Mg-1wt\%Gd alloy during directional solidification under a constant temperature gradient $G$ and constant cooling rate, or equivalent pulling (i.e. isotherms) velocity $V$. 
This velocity is set to 166~\textmu m/s to match the estimated velocity of the isotherms in the experiment (Sec.~\ref{sec:meth:exp}) and we scan a broad range of temperature gradients from 0.3 to 10~K/mm.

Alloy and simulation parameters are summarized in Table~\ref{tab:Param}.
Phase diagram features (i.e. temperatures, concentrations, and resulting $k$ and $m$) for a dilute Mg-Gd alloy were taken from ASM handbooks \cite{okamoto1990binary,valencia2008asm}.
The solid-liquid interface excess free energy, $\gamma_0=89.9~$mJ/m$^2$, and its anisotropy coefficients, $\varepsilon_{66}\approx0.003$ and $\varepsilon_{20}\approx-0.026$, come from molecular dynamics calculations for pure Mg \cite{sun06}.
The Gibbs-Thomson coefficient was also calculated for pure Mg as $\Gamma=\gamma_0T_M/(\rho L_f)$, with a density $\rho=1.59\times10^6~$g/m$^3$ and a latent heat of fusion $L_f=349~$J/g \cite{valencia2008asm}.
The diffusion coefficient for Gd in liquid Mg was approximated as $D\approx D_{\rm Mg}r_{\rm Gd}/r_{\rm Mg}$ \cite{roy88}, considering Mg self-diffusion coefficient $D_{\rm Mg}\approx6.3\times10^{-9}~$m$^2$/s at $T=1000~$K \cite{wang13} and atomic radii $r_{\rm Gd}=233$~pm and $r_{\rm Mg}=145$~pm \cite{periodictable}.

\begin{table*}[htb]
\centering
\caption{\label{tab:Param}
Alloy, process, and simulation parameters.
}
\hskip-50pt
\begin{tabular}{lcrl}
\\
Parameter & Symbol & Value & Unit \\
\hline
\hline
Alloy Gd concentration&$c_\infty$&$1.0$&wt\% Gd \\
\hline
Solute partition coefficient&$k$&$0.4$&-  \\
\hline
Liquidus slope&$m$&$-1.66$&K/wt\% Gd \\
\hline
Pure solvent (Mg) melting temperature &$T_M$&$923$&K  \\
\hline
Alloy liquidus temperature& $T_L$&$921.34$&K  \\
\hline
Alloy solidus temperature&$T_0$&$918.85 $&K  \\
\hline
Anisotropy strength in the basal plane \cite{sun06} & $\varepsilon_{66}$&$0.003$&$-$\\
\hline
Liquid Gd diffusion coefficient \cite{wang13,roy88,periodictable}& $D$&$3.9\times10^{-9}$&m$^2$/s \\
\hline
Gibbs-Thomson coefficient \cite{valencia2008asm,sun06} &$\Gamma$&$1.5 \times 10^{-7} $&K.m  \\
\hline
Solute capillary length at $T_0$&$d_0$&$0.06$& \textmu m \\
\hline
Pulling (isotherms) velocity & $V$&166&\textmu m/s \\
\hline
Thermal gradient & $G$ &$0.3 - 10.0$&K/mm \\
\hline
Diffuse interface width & $W$ & $ 24.74$& $d_0$ \\
\hline
Grid step size & $\Delta x$ & $1.0$& $W$ \\
&&$=1.5$&\textmu m \\
\hline
Noise amplitude \cite{karrap99,ghmadh14} &$\eta$&0.01&$-$ \\
\hline
\end{tabular}
\end{table*}

The initial configuration of the system is close to that used in previously reported phase-field studies of grain growth competition \cite{tourretkarma15,tourret17}.
Two grains A and B are placed on the left (cold) side of a two-dimensional (2D) domain.
We use the notation A and B only to refer to two grains within a simulation (see, e.g., bottom of Fig.~\ref{Fig:Imp}), while we use numbers 1 and 2 to denote grains for each individual GB (top of Fig.~\ref{Fig:Imp}).
In the latter, by convention, grain 1 is above and grain 2 is below the GB if the growth direction goes left-to-right.
Hence, in each simulation containing two GBs, grains A and B alternatively play the role of grains 1 and 2 (or vice-versa) depending on which GB is analyzed.

The nearly-flat solid-liquid interface is initialized at a given undercooling ($\Delta=0.15$) located at $x=L_{\rm solid}$ in the simulation domain.
Grains A and B initially share equally the solid domain, with GBs located along the $(y=L_y/4)$ and $(y=3L_y/4)$ lines.
In order to break the one-dimensional symmetry of the system, the solid-liquid interface location has a small sinusoidal perturbation of amplitude $\Delta x$ and 5 periods (i.e. bumps) across the $y$ dimension.
The preconditioned phase field $\psi$ is initialized as a signed distance function from the interface location, with $\psi>0$ in the solid ($x<L_{\rm solid}$) and $\psi<0$ in the liquid ($x>L_{\rm solid}$).
The concentration field, $c$, is initialized with equilibrium concentrations on both sides of the interface, homogeneous $c$ in the solid, and in the liquid an exponential profile $c\sim\exp\{xV/D\}$ decaying toward $c_\infty$ when $x\to\infty$. 
The $c$ ($\langle 0001 \rangle$) axis of both grains is perpendicular to the simulation plane, and their $a$ ($\langle 10\bar10 \rangle$) axis makes an angle $\alpha_{A}$ or $\alpha_{B}$, respectively, with the $G$ direction, i.e. with the $x$ axis (see Fig.~\ref{Fig:Imp}). 
We use an auxiliary integer field to track the grain identity ($-1$ in grain A, $+1$ in grain B, and $0$ in the liquid).
The grain index field is propagated using the sign of the average over the eight neighboring points whenever a liquid point reaches $(1-\varphi)^2>0.01$ \cite{tourretkarma15,tourret17}.

Simulation sizes are summarized in Table~\ref{tab:Grid}.
The domain length $L_x$ is chosen long enough (i) to track the emergence of tertiary branches on the solid side and (ii) to allow the solute field to relax to $c_\infty$ on the liquid side. 
The domain width $L_y$ is chosen long enough to host at least ten PDAS. 
Hence, $L_y$ is higher for low $G$. 
No-flux and periodic boundary conditions are imposed along $x$ and $y$ directions, respectively. 
The most advanced interface location in the $x$ direction is maintained at a distance $L_{\rm solid}$ of the leftmost ($x-$) domain boundary as listed in Table~\ref{tab:Grid}.

\begin{table*}[htb!]
\centering
\caption{\label{tab:Grid}
Physical dimensions of the system for different thermal gradients and corresponding numbers of mesh points
in the numerical domain.
}
\begin{tabular}{ccccc}
\\
$G$ (K/mm) & $N_{x}\times N_{y} $ & $L_{x}\times L_{y} $ (${\rm mm}^2$) &  $L_{\rm solid}$ (\textmu m) & Time (s)\\ 
\hline
\hline
0.3, 0.5, 1.0, 2.0, 3.0 & $ 960 \times 2560 $ & $ 1.4 \times 3.8    $ & 1150 & 250 \\
\hline
10.0 & $ 576  \times 1024 $ & $ 0.9 \times 1.5 $ & 600  & 80 \\
\hline
\hline
\end{tabular}
\end{table*}

This configuration, illustrated in Fig.~\ref{Fig:Imp}, allows tracking the evolution of one converging GB and one diverging GB for each simulation.
At each GB, the upper ($y+$) and lower ($y-$) grains are labelled as grain 1 and grain 2, respectively. 
At the beginning of each simulation, we impose internal walls (with a no-flux condition in the $y$-direction on both sides) separating the grains in order to discard early transient interactions and thus focus on grain growth competition at a steady velocity $V$ (see Supplementary Materials of \cite{tourret17}). 
Once both grains reach a steady tip undercooling, we gradually remove the ``wall'' separating the two grains and start recording the GB location. 
Simulation time (Table~\ref{tab:Grid}) is selected long enough to track at least ten branching (diverging GB) and ten elimination (converging GB) events or until one grain is eliminated. 
As the moving domain follows the most advanced primary tips, we store the coordinates of the two GBs between grains A and B at the grid points leaving the domain on the $x-$ side. 
The average orientation $\theta$ of each GB is thus extracted via a linear least-square fitting of the GB location points, automated using gnuplot scripts. 
The lateral ($y$) mobility of the GB, or its roughness, is thus represented by the standard deviation of the linear fit.

We aim to scan the full range of bi-crystal configurations $(\alpha_1,\alpha_2)$ for different temperature gradients.
Accounting for the crystal sixfold symmetry in the basal plane, we scan all combinations ($\alpha_A,\alpha_B$) between $-30^\circ$ and $+30^\circ$ with steps of $5^\circ$ for six values of $G$ from 0.3 to 10~K/mm with $V=166~$\textmu m/s (see Fig.~\ref{Fig:Imp} and Table~\ref{tab:Grid}). 
To account for the stochasticity of GB orientation selection, in particular with respect to side-branching, we perform ten simulations for each configuration using different random number seeds for $\beta_{i,j}$ in Eq.~\eqref{eq:noise}.
Simulations with $\alpha_1=\alpha_2$ were only performed once, and primarily used to extract the growth orientation $\alpha$ for a given crystal orientation (in such case, $\alpha\approx\theta$ is directly obtained via the same least-square fitting) as well as the primary spacing.
Simulation results are gathered in maps of the GB angle $\theta$ as a function of $(\alpha_1,\alpha_2)$ similar to those represented for FOG and GL maps in Fig.~\ref{Fig:FOG-GL_abs}. 
Taking advantage of symmetries, one can construct the entire map by scanning about one quarter of the $(\alpha_1,\alpha_2)$ space (also see Supplementary Material of \cite{tourret17}).

As summarized in Table~\ref{tab:configs}, for each $G$, the resulting map corresponds to 10 runs for 42 $(\alpha_A,\alpha_B)$ configurations with $\alpha_A\neq\alpha_B$, while the six configurations with $\alpha_A=\alpha_B$ were performed only once.
The full mapping for the six explored $G$ thus amounts to $6\times426=2556$ simulations.
Each simulation was performed on one Nvidia GPU (GeForce RTX\,3090, RTX\,2080Ti, or Titan\,Xp).
The average compute time for one simulation was about 62~minutes, with individual simulations completed between 6.5~minutes (in the case of quick elimination of a grain) and 3.5~hours (for low $G$ and, typically, $\alpha_1\approx\alpha_2$).  
The cumulative simulation wall time of all single-GPU-parallelized simulations reported here thus amounts to over 2640~hours (i.e., 110~days).

\begin{table}[t!]
\centering
\caption{\label{tab:configs}
Number of simulations performed for each $(\alpha_A,\alpha_B)$ configuration.
}
\begin{tabular}{c | ccccccc}
& \multicolumn{7}{c}{ $\alpha_B (^\circ)$ } \\
$\alpha_A (^\circ)$ & 30 & 25 & 20 & 15 & 10 & 5 & 0 \\
\hline
$-30$ & 10 & \\
$-25$ & 10 &10 & \\
$-20$ & 10 &10 &10 & \\
$-15$ & 10 &10 &10 &10 & \\
$-10$ & 10 &10 &10 &10 &10 & \\
$-5$ & 10 &10 &10 &10 &10 &10 & \\
0 & 10 & 10 & 10 & 10 & 10 & 10 & 1 \\
5 & 10 & 10 & 10 & 10 & 10 & 1 \\
10 & 10 & 10 & 10 & 10 & 1 \\
15 & 10 & 10 & 10 & 1 \\
20 & 10 & 10 & 1 \\
25 & 10 & 1 \\
\end{tabular}
\end{table}

\subsection{Experiments}
\label{sec:meth:exp}

While the article is mainly focused on simulation results, we also performed experiments in order to discuss, specifically, the selection of GB orientation during directional solidification of a Mg-1\,wt$\%$Gd alloy.
We built a custom furnace for directional solidification of thin samples. 
The furnace was designed to later allow in-situ X-ray radiography imaging, but the experiments discussed here were completed without in-situ imaging, i.e. analyses were performed ex-situ after directional solidification. 

\begin{figure}[t!]
\includegraphics[width=3in]{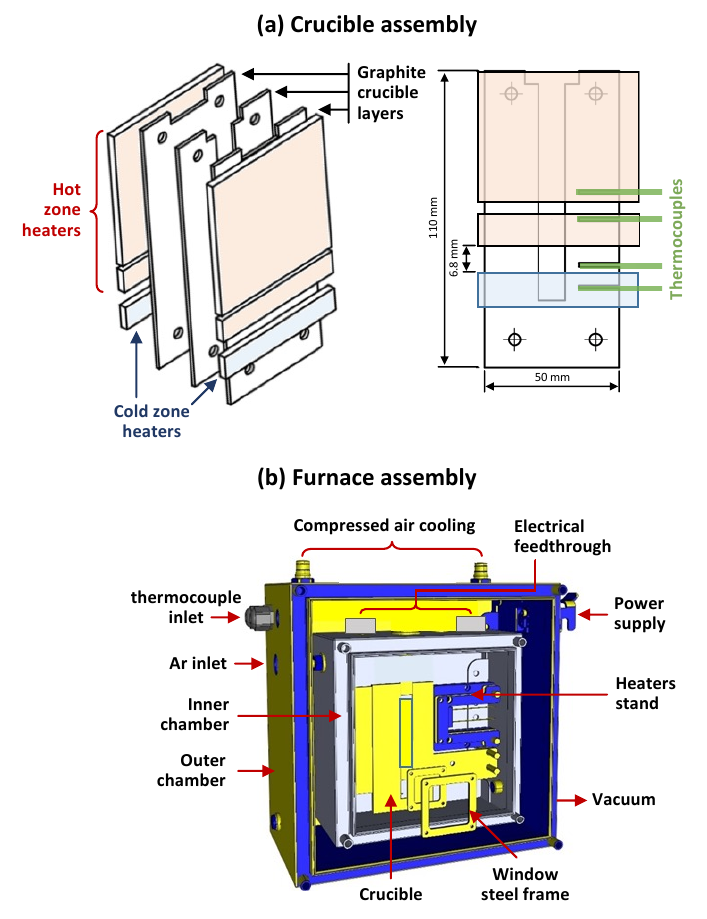}
\caption{
Experimental setup for directional solidification of thin sample under Ar gas atmosphere.
}
\label{Fig:Furnace}
\end{figure}

The furnace, illustrated in Fig.~\ref{Fig:Furnace}, is composed of pairs of identical heaters for ``cold'' and ``hot'' zones separated by 6.8~mm from each other. 
Between the heaters, we insert a graphite crucible consisting of two rectangular plate-like halves and a third one with a central rectangular deep pocket (Fig.~\ref{Fig:Furnace}a). 
A sample of thickness 1~mm is covered in boron nitride, placed into the pocket, and sandwiched between the two crucible plates. 
The temperature field is monitored by four thermocouples located at different heights and embedded in the central layer of the crucible.
The resulting temperature measurements enable the feedback control of the independently controlled flat heaters, as well as the evaluation of the thermal conditions during the solidification experiment. 

As illustrated in Fig.~\ref{Fig:Furnace}b, the system (crucible + sample + heaters + thermocouples) is inserted into an outer chamber. 
Compressed air is flowing between the two chambers, providing thermal insulation and minimizing damage to electrical connections. 
Special feedthrough assemblies were designed to provide an efficient seal on the wires passing through the inner chamber and to restrain them from moving under pressure and vacuum.

We performed directional solidification experiments on thin samples ($15\times 10 \times 1$ mm$^3$) of Mg-1\,wt$\%$Gd. 
Each experiment starts by setting an initial thermal gradient $G$, heating the sample until it melts completely, then cooling down both heaters at a constant rate $R$ while maintaining the thermal gradient $G$ during the process. 
We aimed for a value of $G\approx3~$K/mm and explored cooling rates ranging from 2 to 40~K/min 
Here, we specifically focus on the analysis of one sample that exhibited a few GBs involving grains with a low misorientation of their $c$ axis and the normal direction to the sample plane (hence representative of the 2D simulation configuration).
The thermocouple readings indicate that the sample was solidified under a temperature gradient $G\approx3~$K/mm and a velocity of isotherms $V\approx166~$\textmu m/s.

Electropolishing was performed on the solidified specimen after gently grinding with 1200 grit size abrasive SiC paper. 
The surface of the sample was electropolished at $-30^\circ$C with a voltage of 50~V using a solution of 15~vol$\%$ nitric acid and 85~vol$\%$ ethanol for 30~s. 
The electropolished sample was then cleaned by immersion in an ultrasound bath of pure ethanol for 2 minutes. 
The resulting sample was smooth enough for microstructural characterization in a scanning electron microscope (SEM).

Electron backscatter diffraction (EBSD) was performed to obtain crystallographic orientation maps using a ﬁeld-emission gun SEM (Helios Nanolab 600i) equipped with an Oxford-HKL EBSD detector.
The scanning step size was selected to obtain sufficient spatial resolution to identify GBs. 
The accelerating voltage was set to 20~kV and the current to 2.7~nA. 
An automated mapping was performed that allowed joining several EBSD maps with an imposed overlap (usually 10-20$\%$) to obtain a wide EBSD map. 
The diffraction data was post-processed using Matlab toolbox MTEX \cite{bachmann11}. 
The maps were cleaned using a spline ﬁlter to ﬁll non-indexed points with the nearby pixels and eliminate measurement errors. 
The threshold misorientation angle for grain reconstruction was set to 2°.

The EBSD map was then smoothed and individual GBs were studied in order to select regions of interest for further analysis.
In particular, we focused on regions containing GBs between grains whose $c$ axis deviation from the $z$ axis (noted $\Phi$) is lower than $25^\circ$ (filtered using Matlab MTEX toolbox).
Such GBs are the ones most comparable with our 2D simulations with sixfold anisotropy within the basal plane.
Each GB orientation was then calculated by fitting the location of GB points to straight lines with gnuplot using the same convention as in Fig.~\ref{Fig:Imp}.

\section{Results and discussion}
\label{Sec:result}

\subsection{Differences between cubic and hexagonal systems}
\label{sec:cubic_vs_hexa}

\begin{figure}[t!]
\includegraphics[width=3in]{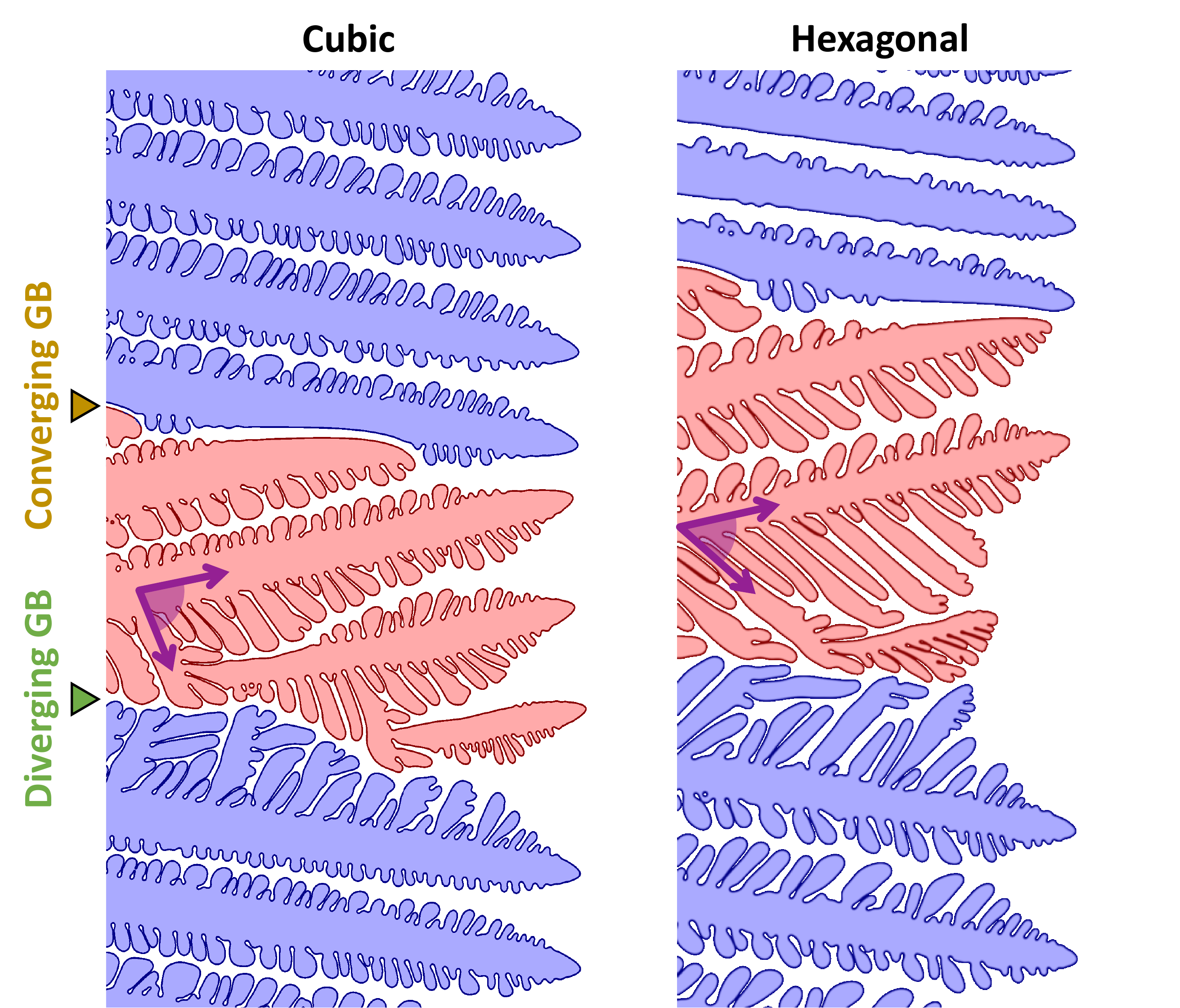}
\caption{
Illustration of grain growth competition mechanisms at converging (top) and diverging (bottom) GBs, highlighting differences between cubic (left) and hexagonal (right) alloy systems.
The left-hand-side is one of the simulations analyzed in Ref.~\cite{tourret17}, corresponding to a succinonitrile-0.4 wt\%acetone alloy at $G=30$~K/cm and $V=25$~\textmu m/s, using simulation parameters from Refs~\cite{tourretkarma15,tourret17}.
The right-hand-side simulation is one of the simulations analyzed here.
Both simulations have $\alpha_A=-10^\circ$ (blue grain) and $\alpha_B=+15^\circ$ (red grain).
}
\label{Fig:CubicVsHexa}
\end{figure}

Before presenting a quantitative analysis of our results, we highlight some of the main differences expected between a cubic system, such as those analyzed in previous articles \cite{tourretkarma15,tourret17,dorari22}, and the hcp alloy system studied here.
These differences are illustrated in Fig.~\ref{Fig:CubicVsHexa}, comparing one of the simulations from Ref.~\cite{tourret17} and one of the current simulations, both of them with $\alpha_A=-10^\circ$ (blue grain) and $\alpha_B=+15^\circ$ (red grain).

As mentioned in Sec.~\ref{sec:meth:simu}, a first difference is the range of angles that needs to be mapped, namely $[-45^\circ;+45^\circ]$ for cubic systems and $[-30^\circ;+30^\circ]$ for hexagonal ones. 
In terms of growth competition, the sixfold symmetry does not lead to a substantial difference at converging GBs (top GBs in Fig.~\ref{Fig:CubicVsHexa}) controlled by dendrite impingement.
In contrast, diverging GBs (bottom GBs in Fig.~\ref{Fig:CubicVsHexa}), piloted by side-branching competition, are expected to exhibit more prominent differences due to the distinct angle between primary dendrites and higher-order side-branches (illustrated by purple arrows in red grains of Fig.~\ref{Fig:CubicVsHexa}).
On the one hand, in cubic (fourfold) systems, the angle between primary and secondary branches is $\approx90^\circ$.
As a result, for low misorientation angles, side-branches face a relatively small temperature gradient in their respective growth direction.
In the extreme case of a growth direction of primary dendrites oriented along the temperature gradient, secondary branches essentially grow along isotherms.
On the other hand, in hexagonal (sixfold) systems, secondary branches grow at an angle $\approx60^\circ$ with respect to the primary dendrites.
Consequently, even at low misorientation angle, side-branches face a relatively higher temperature gradient in their respective growth direction, which is expected to lead to increased side-branching activity compared to cubic systems.

Note that the angle between the growth directions of primary and secondary branches is most often not exactly $90^\circ$ or $60^\circ$, due to the deviation from crystal orientation ($\alpha\neq\alpha_0$) described in Sec.~\ref{sec:theory:dgp} and discussed in Sec.~\ref{sec:pdas}.

\subsection{Primary dendrite arm spacing}
\label{sec:pdas}

\begin{figure}[t!]
\includegraphics[width=3in]{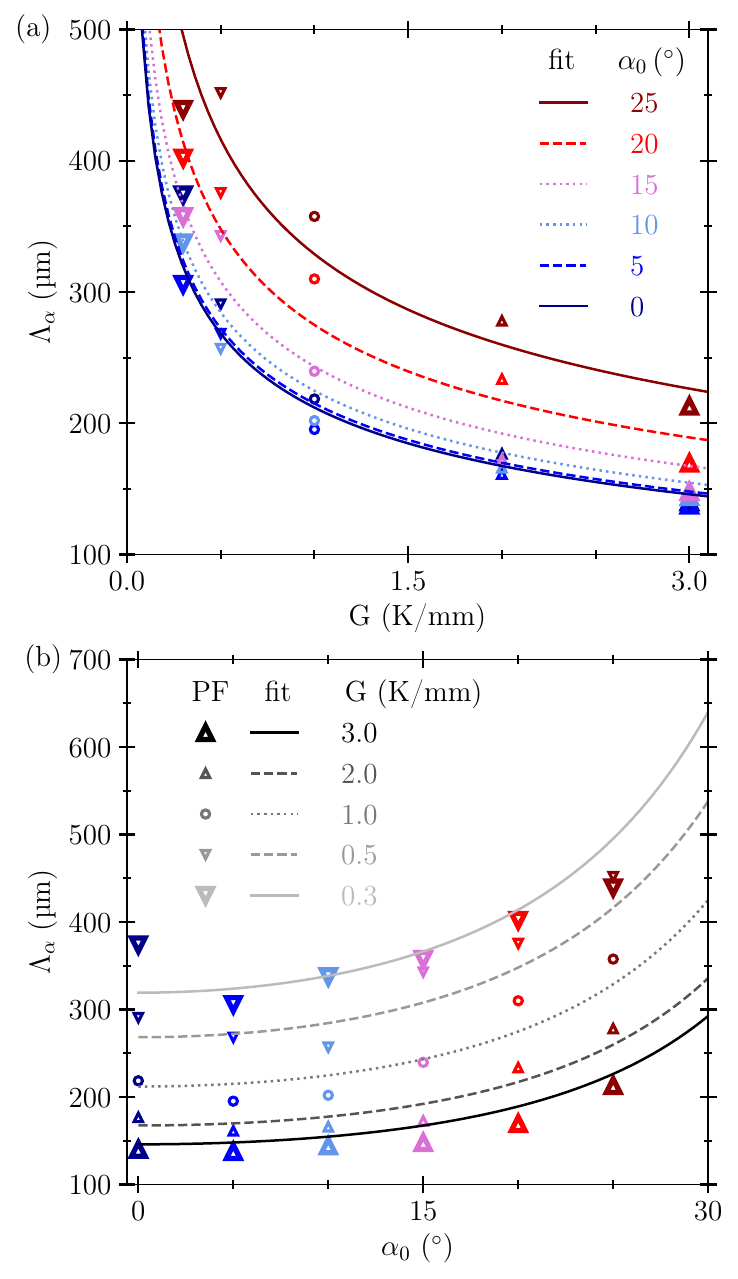}
\caption{
Primary arm spacing ($\Lambda_\alpha$) as a function of (a) the thermal gradient $G$ for different crystal orientations $\alpha_0$, and (b) $\alpha_0$ for different $G$ values, showing PF results (symbols) against fitted Eq.\,\eqref{Eq:GandHcp} (lines) with $L= 0.212\,$mm$^{0.66}$K$^{0.34}$, $c= 0.34$, $d = 0.55$ and  $e= 3.0$. 
In both plots, colors represent the crystal orientation $\alpha_0$ from $0^\circ$ (dark blue) to $25^\circ$ (dark red) while symbols represent temperature gradient $G$ from 0.3~K/mm (big $\bigtriangledown$) to 3.0~K/mm (big $\bigtriangleup$).
}
\label{Fig:LambdaG}
\end{figure}

Fig. \ref{Fig:LambdaG} shows the variation of selected PDAS $\Lambda_\alpha$ in PF simulations (symbols) as a function of (a) $G$ and (b) $\alpha_0$ together with a set of fitted curves (lines) using a slightly modified version of Eq.~\eqref{Eq:Gand}: 

\begin{equation}
\Lambda_\alpha(G,\alpha_0)=LG^{-c}\{1+d[\cos({45 \alpha_0}/{30} )^{-e}-1]\}
\label{Eq:GandHcp}
\end{equation}
that includes an empirical correction factor $45/30$ in the cosine function to account for the sixfold symmetry of the hcp crystal. 
For simplicity, we express $\Lambda_\alpha$ directly as a function of the crystal orientation $\alpha_0$, rather than the dendrite growth direction $\alpha$, considering that this will likely only affect the value of the fitted coefficients but not the validity of the relation.
All curves correspond to Eq.~\eqref{Eq:GandHcp} with $L\approx 0.212\,$mm$^{0.66}$K$^{0.34}$, $c\approx 0.34$, $d \approx 0.55$ and  $e\approx 3.0$, obtained from a global fit of all data points for $0.3\leq G\leq3.0$~K/mm.
The highest gradient, $G=10$~K/mm, was excluded from the fit on purpose, as it leads to intermediate patterns between dendritic and cellular (see Sec.~\ref{sec:microstructures}), hence not entirely relevant to the $\Lambda_\alpha(G,\alpha_0)$ law derived specifically for dendritic arrays.

The overall variation of $\Lambda_{\alpha}$ with $G$ and $\alpha_0$ is in good agreement with previous works in cubic systems \cite{gandin96,tourretkarma15}. 
Indeed, spacings decrease with increasing $G$ with $\Lambda\sim G^{-0.34}$ (Fig.~\ref{Fig:LambdaG}a) and they increase with increasing $\alpha_0$ (Fig.~\ref{Fig:LambdaG}b).
Due to the low number of measured spacings (and hence limited statistical significance), we did not perform a fully quantitative analysis of spacing distributions, nor of the full PDAS stability limits, as, e.g., in \cite{clarke2017microstructure, song2018thermal, bellon21} (out of scope here).
However, our observations suggest that the distribution of spacing is wider for lower values of $G$.
This supports our earlier results on Al-Cu alloys \cite{bellon21}, which showed that the spacing stability range $[\Lambda_{\rm min},\Lambda_{\rm max}]$ becomes wider as $G$ decreases (in the dendritic regime) and that $\Lambda_{\rm max}/\Lambda_{\rm min}\to2$ as $G$ increases (toward the cellular regime).
Overall, Fig.~\ref{Fig:LambdaG} illustrates that $\Lambda_\alpha$ seems relatively weakly affected by $\alpha_0$ for moderate misorientations (at most 25 $\%$ for $\alpha_0\leq 25^\circ$) in comparison to its strong dependence upon $G$.  

The effect of $G$ observed here appears weaker compared to previous results for cubic-symmetry alloys \cite{hunt79,Kurz81,gandin96,bellon21}.
Indeed, at a given $\alpha_0$, varying $G$ by an order of magnitude induces a variation of $\Lambda_\alpha $ (i.e., Pe) by a factor of $10^{0.34}\approx 2.2$, while it typically does so by a factor of $10^{0.5}\approx 3.2$ for alloys with a cubic symmetry.
This weaker dependence upon $G$ might be attributed to the fact that secondary arms, growing at an angle of 60$^\circ$ from the primary trunks, face a higher temperature gradient in their growth direction than in cubic systems, for which secondary arms might even grow along isotherms when $\alpha_0=0^\circ$.
In the case of hcp systems, the interactions between dendrites through the solute field might then be more intense than in cubic systems, and the emergence of tertiary branches as new primary dendrites might also occur earlier (i.e. at lower $\Lambda$), resulting in a PDAS stability domain narrower in hcp systems.  
While PDAS selection is not directly the main focus of the present article, it might be of interest to address these interpretations in future studies, i.e. by comparing PDAS stability domains for model alloys with similar thermophysical parameters but different crystalline symmetries.

\subsection{Dendritic growth direction} 
\label{Sec:DGP}

\begin{figure}[t!]
\centering
\includegraphics[width=3in]{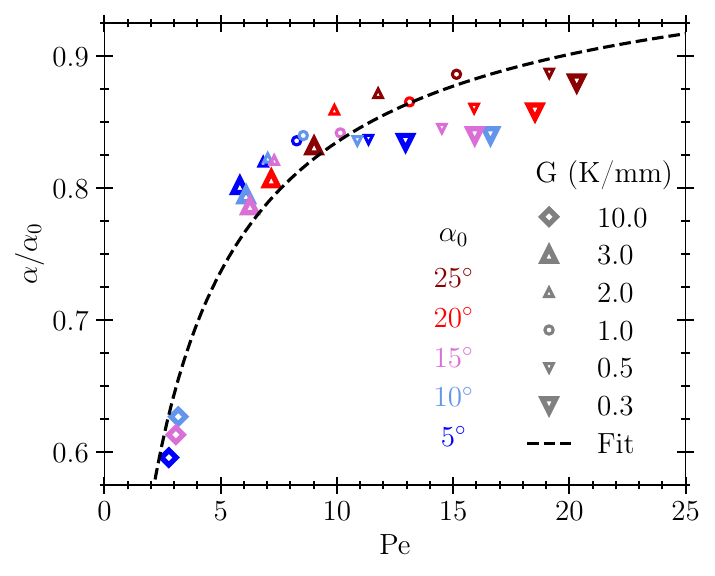}
\caption{
Dendrite growth orientation (i.e. ratio of dendrite growth direction angle $\alpha$ by crystal orientation $\alpha_0$)  versus P\'eclet number Pe$=\Lambda_0V/D$, compared to fitted Eq.\,\eqref{Eq:DGP} with $f = 0.708$ and $g = 0.854$ (dashed line). 
Consistently with Fig.\,\ref{Fig:LambdaG}, different symbols represent different temperature gradients, $G$, while colors stand for crystal orientations, $\alpha_0$ .
}
\label{Fig:DGP}
\end{figure}

Fig.~\ref{Fig:DGP} shows the ratio $\alpha/\alpha_0$ between growth direction, $\alpha$, and crystal orientation, $\alpha_0$, measured in the PF simulations as a function of the spacing P\'eclet number, ${\rm Pe} =\Lambda_0 V/D$, for $G=[0.3-10.0]$~K/mm (symbols) and a least-square fit (dashed black line) of Eq.~\eqref{Eq:DGP}.

Overall, our results show that the dependence of growth direction established for cubic systems also applies for Mg-1wt\%Gd (hcp system), with all data points relatively well fitted by Eq.~\eqref{Eq:DGP}, here for $f = 0.708$ and $g = 0.854$. 
Similarly as in cubic systems \cite{akamatsu97,dgp08,ghmadh14,tourretkarma15,dorari22}, $\alpha\to\alpha_0$ as Pe increases, i.e. as $G$ decreases, with a microstructure pattern within the well-developed dendritic regime at low $G$ (i.e. high Pe).
Approximating $f(\alpha_0)$ by a constant value also seems to remain a valid assumption. 
Fitted parameters, however, differ from those assessed in cubic systems (e.g. $f\approx 0.2$ and $g\approx 1.7$ in   \cite{tourretkarma15}).

For a given temperature gradient, values of the ratio $\alpha/\alpha_0$ fall within a narrow range, as summarized in Table~\ref{tab:DGP}.
Notably, all temperature gradients exhibit a ratio $\alpha/\alpha_0$ higher than 0.78 in all simulations/configurations, except for the highest $G$ at 10~K/mm.
This is a clear indication of the latter belonging to a transition regime between dendrites ($\alpha\approx\alpha_0$) and cells ($\alpha\approx0$), while all $G\leq3~$K/mm cases are within the dendritic regime.
As patterns transition toward the cellular regime, microstructures actually exhibit increasingly shallower cells and transition to a seaweed pattern as $\alpha_0$ increases, thus preventing the identification and measurement of clear primary spacings (hence only reported for $\alpha_0\leq15^\circ$ at $G=10~$K/mm in Fig.~\ref{Fig:DGP}).
Moreover, the overall grouping of data points for $G\leq3~$K/mm may also be related to the weaker dependence of $\Lambda_0$ on $G$ than in cubic systems, mentioned in Sec.~\ref{sec:pdas}, resulting in a narrower range of Pe numbers.

\begin{table}[t!]
\centering
\caption{\label{tab:DGP}
Ratio between growth angle $\alpha$ and crystal orientation angle $\alpha_0$ for each temperature gradients $G$.\\
}
\begin{tabular}{c|ccc}
$G$ & \multicolumn{3}{c}{ $\alpha/\alpha_0$ } \\
(K/mm) & Min & Max & Mean \\
\hline
\hline
0.3 & 0.835 & 0.880 & 0.851 \\
0.5 & 0.836 & 0.887 & 0.853 \\
1.0 & 0.836 & 0.886 & 0.854 \\
2.0 & 0.820 & 0.872 & 0.839 \\
3.0 & 0.787 & 0.832 & 0.805 \\
10.0 & 0.596 & 0.627 & 0.612 \\
\hline
\hline
\end{tabular}
\end{table}

\subsection{Microstructure overall morphologies}
\label{sec:microstructures}

\begin{figure*}[t!]
\includegraphics[width=6.5in]{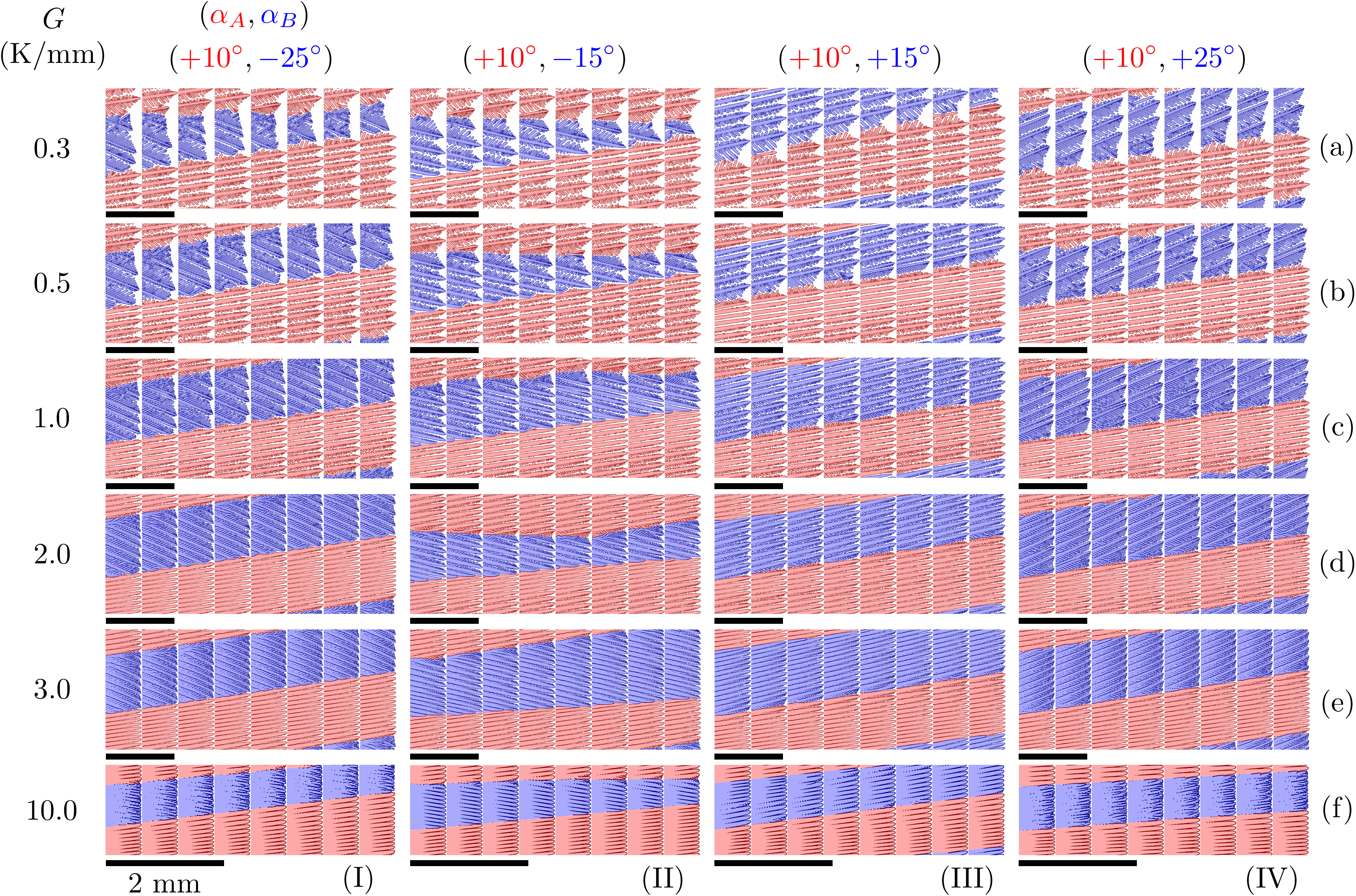}
\caption{
Spatiotemporal reconstruction of PF-simulated microstructures during directional solidification of Mg-1wt\%Gd at $V=166~$\textmu m/s for different temperature gradients $G$ (rows a to f) and crystal orientations $(\alpha_{\rm A},\alpha_{\rm B})$ (columns I to IV).
The total simulated length is typically three to four times longer than that illustrated here.
}
\label{Fig:microstructures}
\end{figure*}

Fig.~\ref{Fig:microstructures} shows reconstructed microstructures (i.e. spatiotemporal snapshots) resulting from typical PF simulations for different temperature gradients (rows) from $G=0.3~$K/mm (a) to 10~K/mm (f) and different bi-crystal orientations (columns), with the orientation of the red grain fixed at $\alpha_{\rm A}=+10^\circ$, and that of the blue grain with $\alpha_{\rm B}=-25^\circ$ (I), $-15^\circ$ (II), $+15^\circ$ (III), and $+25^\circ$ (IV).
Fig.~\ref{Fig:microstructures} highlights the well-known difference between converging and diverging grain boundaries, forming via different mechanisms of dendrite impingement or side-branching competition, respectively \cite{tourretkarma15,tourret17}.

The selected orientation $\theta$ of converging GBs ($\alpha_1<\alpha_2$) -- via primary dendrite impingement -- remains, in all cases, close to the growth direction $\alpha$ of the most favorably oriented grain (here, the red grain).
However, some relatively rare events of well-oriented dendrite elimination appear -- see, e.g., three elimination events in the illustrated example for $(\alpha_{\rm A},\alpha_{\rm B})=(+10^\circ,-15^\circ)$ at $G=2~$K/mm (d$_{\rm II}$).
These events, often referred to as ``unusual overgrowth'' in the literature \cite{li12,takaki14}, result from a mechanism already explained via PF simulations in cubic systems, namely a progressive local reduction of spacing (i.e. increase of undercooling) of the well-oriented dendrite immediately adjacent to the GB as misoriented dendrites impinge onto it, until its elimination occurs.
These events result in small deviations of the GB orientation from the favorably-oriented grain growth direction. 

Like in cubic systems, the case of diverging GBs ($\alpha_1>\alpha_2$), driven by the growth competition of side-branches, exhibits a greater complexity and variability.
When both grains grow in the same lateral direction ($\alpha_1\times\alpha_2>0$), as illustrated in columns III-IV of Fig.~\ref{Fig:microstructures}, the GB orientation remains relatively close to that of the favorably-oriented grain.
However, some deviation from it may result from rare events of side-branching from the favorably oriented grain, in particular when $\alpha_1$ and $\alpha_2$ are close at low $G$ (see, e.g., Fig.~\ref{Fig:microstructures}a$_{\rm III}$).
When grains grow in opposite lateral directions ($\alpha_1\times\alpha_2<0$), emerging GBs exhibit a high lateral mobility. 
As a result, the GBs may actually go in any direction, and their shape may deviate significantly from straight lines (see, e.g., Fig.~\ref{Fig:microstructures}d$_{\rm II}$).
Because they grow at a $60^\circ$ angle from the primary dendrites, side-branches always face a significant temperature gradient, compared to fourfold (cubic) systems, where grains at $\alpha_0\approx0^\circ$ generate side-branches growing essentially along isotherms, in particular at low $G$ \cite{tourretkarma15}.
At low $G$, the resulting high side-branching activity contributes to the increase in primary spacing, and hence to the available space between grains for the side-branching competition to occur, thus also promoting the GB lateral mobility and the variability of $\theta$ (see Section~\ref{sec:maps}).
At the highest $G=10~$K/mm, microstructures clearly transition toward an intermediate regime between dendritic and cellular, and a crystal orientation $\alpha_0=\pm25^\circ$ even leads to seaweed microstructures (Fig.~\ref{Fig:microstructures}f$_{\rm I}$ and 6f$_{\rm IV}$), not showing any clear primary or secondary branches.

The so-called rare events of elimination (at converging GBs) and of side-branching (at diverging GBs) of well-oriented dendrites are fundamentally different in nature. 
On the one hand, the ``unusual overgrowth'' at converging GBs is directly related to a progressive increase of undercooling of the well-oriented primary dendrite adjacent to the GB, as successive misoriented dendrites impinge upon it at a relatively regular frequency. 
As such, it is expected to have a frequency of occurrence that is directly related to the primary spacings and primary growth directions in both grains, with a relatively narrow statistical variability.
Published results indicate that this frequency tends to decrease when $\Lambda_\alpha$ increase or when $|\alpha_1+\alpha_2|$ increase \cite{tourret17,takaki14}.
On the other hand, unusual events of side-branching from the well-oriented grain are expected to exhibit a much broader level of statistical variability.
This is due to the fact that dendritic side-branching is, by essence, induced by selective amplification of noise, and hence stochastic \cite{pieters1986noise, barber1987dynamics, dougherty1987development, brener1995noise}.
However, in spite of the high number of simulations presented here, the total number of unusual elimination and branching events remains too limited for a significant analysis of their frequency and statistical distribution.
Further investigation remains needed to conclusively establish whether these rare events could be rigorously classified statistically. 

At the scale of an overall grain structure (i.e. averaged over thousands, millions, or more grain GBs), the effect of such small deviations on emerging material properties is also likely small, in particular for converging grain boundaries, which, for the most part, remain relatively close to straight lines (Fig.~\ref{Fig:microstructures}).
However, at the scale of individual GBs, given the intricate relationship between GB structure and morphology with mechanical properties, one may expect that deviations from a straight morphology may influence the properties (e.g. increasing strength by disrupting the transmission of dislocation or twins, or the extent of GB sliding).

\subsection{GB orientation selection maps}
\label{sec:maps}

\begin{figure*}[t!]
\centering 
\includegraphics[width=4.9in]{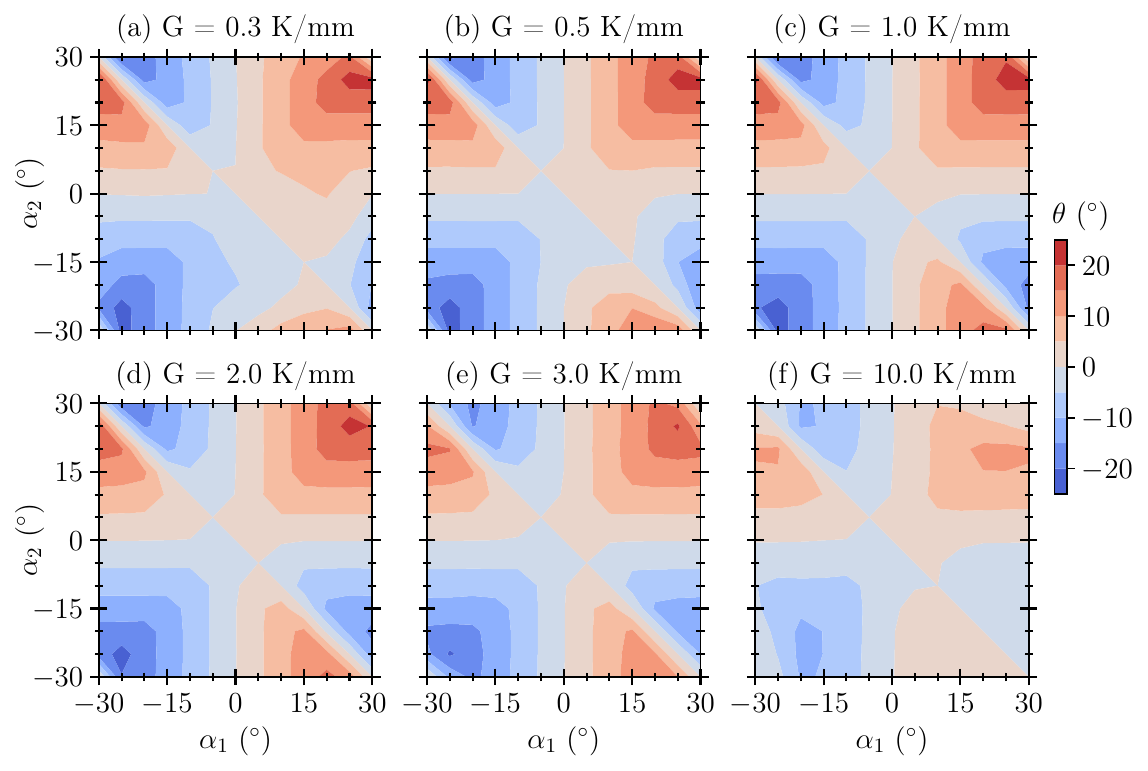}
\caption{
PF-predicted selected GB orientation $\theta$ as a function of the crystal orientation angles ($\alpha_1,\alpha_2$) for different values of the temperature gradient $G$ for $V=166$\,\textmu m/s.
Note that the the color bar range $(-25^\circ,+25^\circ)$ is slightly narrower than in Fig.~\ref{Fig:FOG-GL_abs} $(-30^\circ,+30^\circ)$ due to the deviation of growth orientation from crystal orientation (see Section~\ref{Sec:DGP}) while the theoretical FOG and GL maps (Fig.~\ref{Fig:FOG-GL_abs}) consider $\alpha=\alpha_0$.
}
\label{Fig:maps_PF}
\end{figure*}

\begin{figure*}[t!]
\centering 
\includegraphics[width=4.9in]{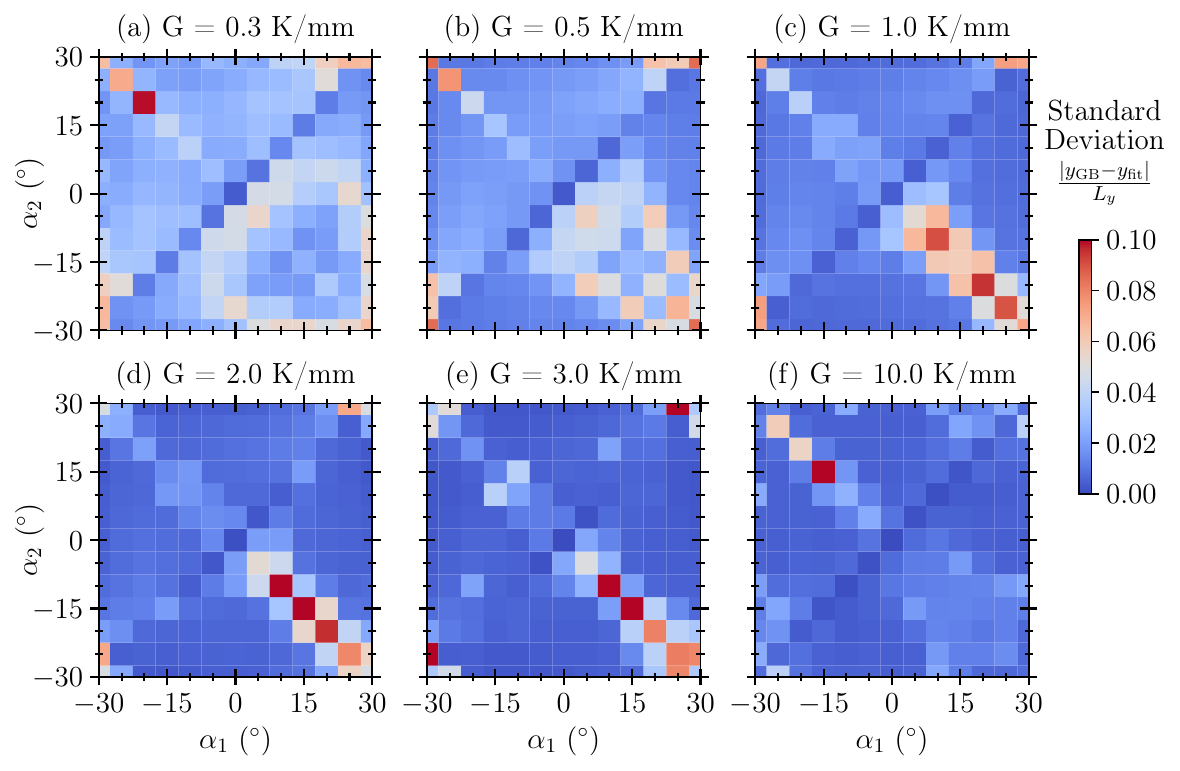}
\caption{
Standard deviation from the least-square fit of the PF-predicted GB trajectories ($y_{\rm GB}$) to a straight line trajectory ($y_{\rm fit}$) normalized by the the domain width ($L_y$).
}
\label{Fig:maps_StdDev}
\end{figure*}

\begin{figure}[t!]
\includegraphics[width=2.8in]{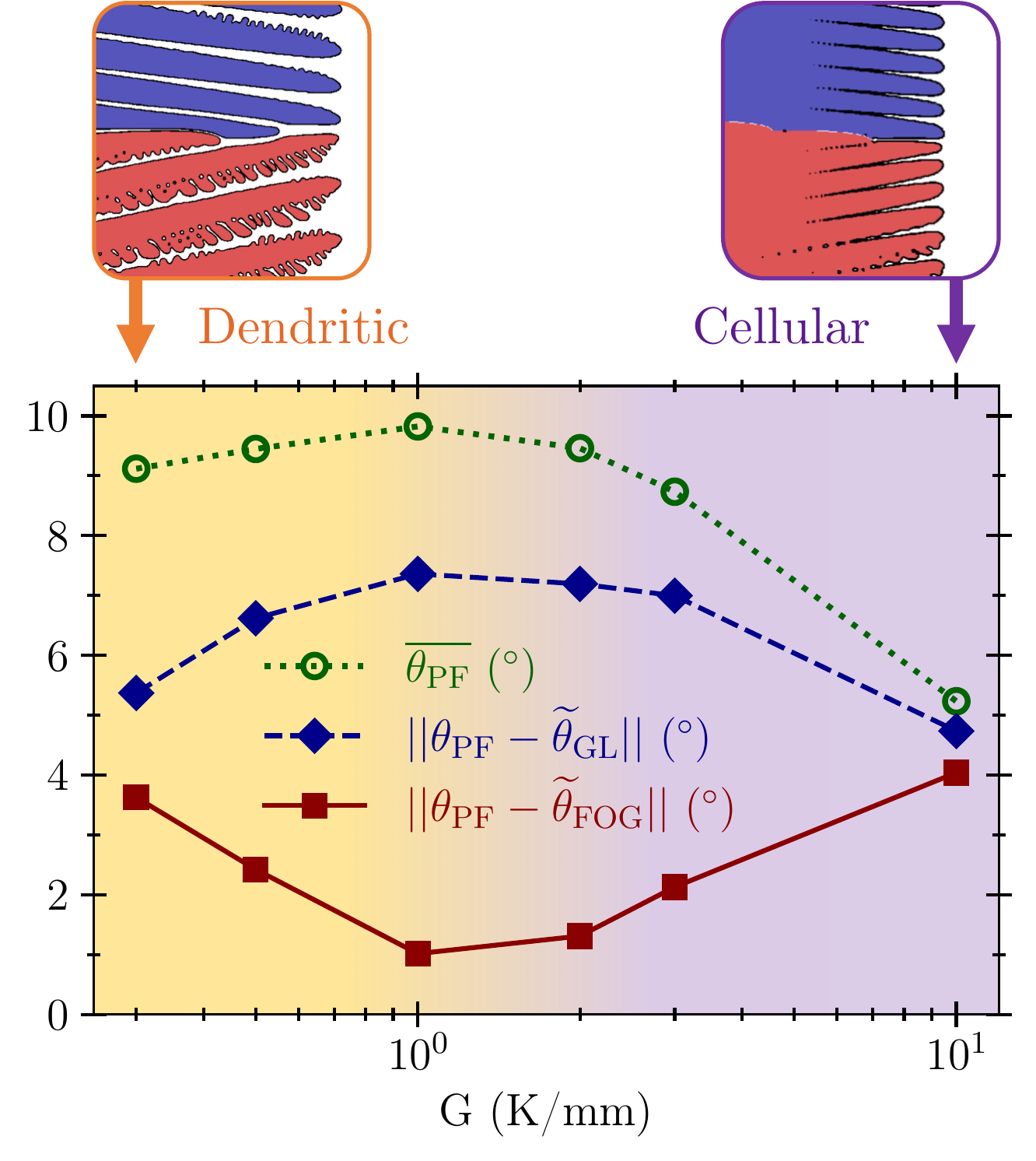}
\caption{
Absolute deviation, averaged over all $(\alpha_1,\alpha_2)$ orientations, of PF-predicted GB orientation, $\theta_{\rm PF}$, compared to the theoretical FOG criterion, $\widetilde\theta_{\rm FOG}$, and to the GL, $\widetilde\theta_{\rm GL}$, both corrected to account for $\alpha/\alpha_0\neq1$ (see text) and average GB orientation of the overall PF map, $\overline{\theta_{\rm {PF}}}$, as a function of the temperature gradient $G$.
The color background represents the transition between dendritic (orange) and intermediate dendritic-cellular (purple) regimes, with illustrative microstructures shown at the top for $G=0.3$ and 10~K/mm.
}
\label{Fig:Sum}
\end{figure}

Fig.~\ref{Fig:maps_PF} shows the GB orientation $\theta$ as a function of ($\alpha_1,\alpha_2$) obtained by phase-field simulations for  $G=[0.3-10]~$K/mm. 
Each data point is an average of ten simulations with different fluctuations, except simulations for $\alpha_1=\alpha_2$ performed only once (see Table~\ref{tab:configs}). 
Importantly, Fig.~\ref{Fig:maps_PF} data points along the $(\alpha_1=-\alpha_2)$ line, as well as ($\alpha_1=\alpha_2=\pm30^\circ$) cases are set to $\theta=0^\circ$, due to symmetry considerations. 
Indeed, for these simulations, a vertical mirror symmetry leads to similar values of $(\alpha_1,\alpha_2)$ -- yet interchanging the roles of grains A and B as grain 1 or 2 -- but an opposite value of $\theta$. 
As such, considering $\theta\neq0^\circ$ for these cases would actually represent an arbitrary choice between these two opposite yet equivalent cases.

Fig.~\ref{Fig:maps_StdDev} shows the corresponding maps of the average standard deviation of the linear fit of the $y-$position of the GB, $y_{\rm fit}$, fitted to the PF-predicted position, $y_{\rm GB}$, normalized by the domain total height $L_y$. 
These maps qualitatively show the nonlinearity (or roughness) of the GB, which also illustrates the lateral ($y$) mobility of the GB as it is forming.

Fig.~\ref{Fig:Sum} shows the deviation of the PF-calculated $\theta$ map over the entire $(\alpha_1,\alpha_2)$ space compared to theoretical predictions using $\alpha/\alpha_0$-corrected FOG (Fig.\,\ref{Fig:FOG-GL_abs}a) and GL (Fig.\,\ref{Fig:FOG-GL_abs}b) models, graphed as a function of the temperature gradient $G$.
Therein, FOG and GL grain boundary orientations, respectively $\theta_{\rm FOG}$ and $\theta_{\rm GL}$, were corrected by a multiplicative factor $\alpha/\alpha_0$ in order to account for the deviation of dendrite growth direction ($\alpha$) from the crystal orientation ($\alpha_0$).
Since $\alpha/\alpha_0$ is significantly dependent on $G$ only, we used a constant value for each $G$ (i.e. the mean value listed in Table~\ref{tab:DGP}).
The resulting corrected maps are denoted $\widetilde\theta_\mu$, with $\mu\in\{{\rm FOG},{\rm GL}\}$, and the deviation with phase-field results is quantified as $||\theta_{\rm PF}- \widetilde\theta_\mu||$ defined as the square root of the average value of $(\theta_{\rm PF}- \widetilde\theta_\mu)^2$ over the $(\alpha_1,\alpha_2)$ space.
Fig.~\ref{Fig:Sum} also shows the PF-predicted value of GB orientation averaged over the entire $(\alpha_1,\alpha_2)$ map ($\overline{\theta_{\rm PF}}$).

Both the GB orientation selection maps (Fig.~\ref{Fig:maps_PF}) and the $||\theta_{\rm PF}- \widetilde\theta_\mu||$ deviation plots (Fig.~\ref{Fig:Sum}) exhibit two distinct trends at low $G$ and at high $G$, which we attribute to a transition from a well-developed dendritic regime ($G\leq1$~K/mm) to an intermediate regime between dendritic and cellular ($G\geq2$~K/mm).
This transition is illustrated with the color gradient background of Fig.~\ref{Fig:Sum}, where illustrated microstructures for the two extreme values of $G$ clearly show the difference between growth morphologies (also see Fig.~\ref{Fig:microstructures}).

Within the dendritic regime ($G\leq1$~K/mm), the $\theta_{\rm PF}$ selection map gets closer to the GL and further from the FOG criterion as $G$ decreases, as the pattern becomes more pronouncedly dendritic, while the overall average GB orientation does not really exhibits any significant variation ($9^\circ<\overline{\theta_{\rm PF}}<10^\circ$).
In Fig.~\ref{Fig:maps_PF}, most of the $\theta(\alpha_1,\alpha_2)$ map exhibits little variation within this $G$ range, but the most striking change appears for diverging GBs with $\alpha_1\times\alpha_2<0$, i.e. the bottom-right quadrant of maps.
This transition from FOG to GL as $G$ decreases in the dendritic regime is consistent with prior observation from PF simulation of fourfold symmetry systems \cite{dorari22}.
We also confirmed it further via simulations at lower $G$ for selected $(\alpha_1,\alpha_2)$ configurations.

\begin{figure*}[t!]
\centering
\includegraphics[width=4.in]{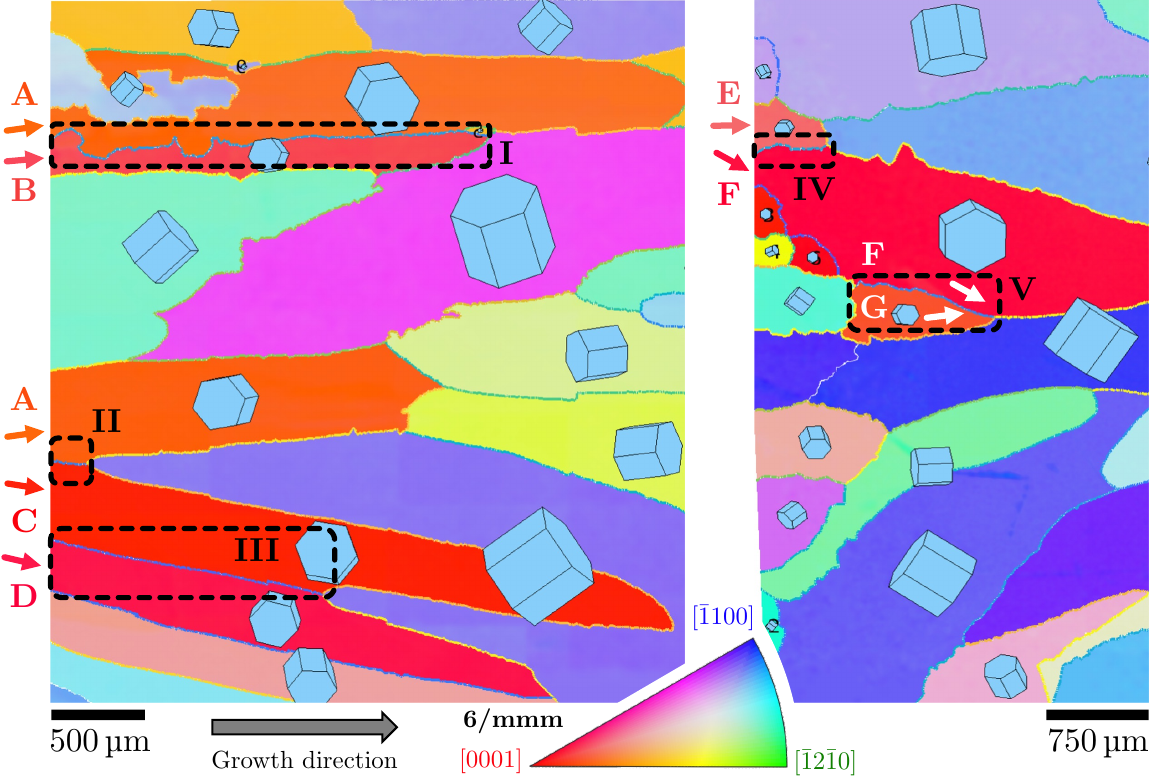}
\caption{
EBSD inverse pole figure maps obtained from two different regions of a Mg-1wt\%Gd sample directionally solidified at a cooling rate $R=30~$K/min and $G=3$~K/mm (i.e. $V\approx166~$\textmu m/s). Grains with relatively low $c-$axis misorientation are labelled A through G and shown next to arrows illustrating the corresponding crystal orientation $\alpha_0$, and resulting grain boundaries are labelled I through V (see Table~\ref{Tab:PFvsExp}).
}
\label{Fig:EBSD}
\end{figure*}

\begin{table*}[t!]
\caption{
Comparison of experimental GB orientations measured in Fig.~\ref{Fig:EBSD} ($\theta_{\rm Exp}$) with PF-predicted ones ($\theta_{\rm PF}$) interpolated from $\theta(\alpha_1,\alpha_2)$ of Fig.~\ref{Fig:maps_PF}e, for selected grain boundaries with relatively low $c-$axis misorientation ($\Phi$) with respect to the normal to the sample/imaging plane. \\
}
\label{Tab:PFvsExp}
\centering
\begin{tabular}{ c c r r r r r r c c}
GB & Grains 1/2 &$\alpha_1$($^\circ$)&$\alpha_2$($^\circ$)&$\theta _{\rm Exp}$($^\circ$)&$\theta _{\rm PF}$($^\circ$)& $\Phi_1$($^\circ$)&$\Phi_2$($^\circ$) & Type & $\alpha_1\times\alpha_2$ \\ 
\hline
\hline
I & A/B & 8.1 & 5.6 & 2 & 4.5 & 15.4 & 21.8 & \color{Purple}{Diverging} & \color{Blue}{$>0$} \\ 
\hline
II & A/C & 8.1 & $-9.5$ & $-4$ & 0.9 & 15.4 & 5.8 & \color{Purple}{Diverging} & \color{Red}{$<0$} \\ 
\hline
III & C/D & $-9.5$ & $-13.6$ & $-12$ & $-8.1$ & 5.8 & 14.5 & \color{Purple}{Diverging} & \color{Blue}{$>0$} \\ 
\hline
IV & E/F & 1.5 & $-29.8$ & $-1.6$ & 1.1 &  20.6 &12.3 & \color{Purple}{Diverging} & \color{Red}{$<0$} \\ 
\hline
V & F/G & $-29.8$  & 7.2 & 16 & 6.2 & 12.3 & 15.5  & \color{Orange}{Converging} & \color{Red}{$<0$} \\ 
\hline
\hline
\end{tabular}
\end{table*}

For high temperature gradients ($G\geq2~$K/mm), the microstructure experiences a transition between dendritic (low $G$) and cellular (high $G$) regimes.
The deviation against FOG and GL (Fig.~\ref{Fig:Sum}) shows a reverse trend compared to the dendritic regime, but rather than strictly approaching the GL at high $G$, this change is principally caused by the significant decrease in $\overline{\theta_{\rm PF}}$ with $G$.
Indeed, as $G$ increases and the microstructure approaches the cellular regime, not only does the growth directions $\alpha_1$ and $\alpha_2$ tend toward the temperature gradient direction, but so does the GB orientation.
This is apparent in Fig.~\ref{Fig:maps_PF} by the significant reduction of $|\theta|$ throughout the entire map at high $G$.
We did not explore temperature gradients much higher than 10~K/mm because, as the cells get shallower and the interface approaches a nearly-planar interface, the dynamics of triple points, not properly accounted for in the current single-phase-field model, starts taking a prominent role in the selection of the GB orientation \cite{ghosh2019influence}.

In terms of GB orientation variability and lateral mobility (Fig.~\ref{Fig:maps_StdDev}), the highest deviations appear systematically for diverging GBs (bottom-right half), in particular for $(\alpha_1\times\alpha_2<0)$ configurations (bottom-right quadrant), as well as at (or close to) symmetric configurations, i.e. for $(\alpha_1\approx-\alpha_2)$.
High variability also appears close to the $\alpha_i=\pm30^\circ$ corners ($i\in\{1,2\}$), in relation to the microstructure transition toward a seaweed pattern.
As mentioned in Sec.~\ref{sec:microstructures} (Fig.~\ref{Fig:microstructures}), the side-branching activity along diverging GBs is promoted by a low $G$.
This results in a broader region in the $(\alpha_1,\alpha_2)$ space that exhibits a significant deviation, compared to higher $G$ where the high variability is concentrated mostly on symmetric $(\alpha_1=-\alpha_2)$ configurations.
Interestingly, at both low and high $G$, a region of high variability also emerges along the $(\alpha_1=-\alpha_2)$ line for converging GBs (top-left half).

\subsection{Comparison with experiments}

Fig.~\ref{Fig:EBSD} shows EBSD maps obtained from two different regions of a Mg-1wt$\%$Gd sample after a directional solidification at a cooling rate $R=30~$K/min and $G=3~$K/mm (i.e. $V=166$~\textmu m/s). 
These regions were selected because they contain GBs between grains whose $c$ axis deviation from the $z$ axis ($\Phi_i$ in Table~\ref{Tab:PFvsExp}) is lower than $25^\circ$.
These grains are labelled A through G, and the five resulting GBs are labelled I through V.
Table~\ref{Tab:PFvsExp} summarizes the grain orientations $(\alpha_1,\alpha_2)$ within the $(x,y)$ plane, the tilt angle $\Phi$ between the $c$-axis of both grains and the $z$ coordinate axis, as well as measured GB orientation, $\theta_{\rm Exp}$, compared to the PF-predicted orientation, $\theta_{\rm PF}$.
The latter is calculated by bilinear interpolation from the $\theta(\alpha_1,\alpha_2)$ map of Fig.~\ref{Fig:maps_PF}e.

The comparison between simulation and experiments, summarized in Table~\ref{Tab:PFvsExp}, show some qualitative agreement, but also highlights some notable discrepancies. While three GB (I, III, V) exhibit orientations of the same sign, two actually have opposite signs (II, IV).
The latter correspond to diverging GBs with $\alpha_1\times\alpha_2<0$, which we have shown to be the most prone to high deviations. 
These GB angles also have at most $4^\circ$ in absolute value, making the change of sign within the expected margin of stochastic variability.
Among the other three GBs, the two orientations closest to PF predictions exhibit small deviations of $2.5^\circ$ (I) and $3.9^\circ$ (III), with both being diverging GBs with $\alpha_1\times\alpha_2>0$, which are expected to show relatively low variability (Sec.~\ref{sec:maps}).
The last GB (V), showing a discrepancy of almost $10^\circ$, is a converging GB with $\alpha_1\times\alpha_2<0$ and, importantly, it involves a grain (F) with an absolute misorientation close to $\pm30^\circ$, putting it within the region of highest variability among converging GBs (see Fig.~\ref{Fig:maps_StdDev}e).
Another factor that could contribute to the discrepancy between simulation and experiments is the relatively small observable length of the GB, in particular for GBs II, IV, and V (Fig.~\ref{Fig:EBSD}).

In addition to the limited statistical sample size, another very important source of possible discrepancy between simulation and experiments is the inherent comparison of three-dimensional experiments with two-dimensional simulations.
Even though we filtered out the grains with a basal plane most titled from the observation plane to focus on GBs with both $\Phi_i<25^\circ$, the tilt angles of analyzed grains ($\Phi_i$ in Table~\ref{Tab:PFvsExp}) are not negligible (from 5.8 to $21.8^\circ$).
Three-dimensional grain growth competition, in addition to resulting in additional degrees of freedom to describe the 3D bi-crystalline orientations, have been shown to result in unexpectedly complex mechanisms \cite{takaki18,song2023cell}.
These not only include GB orientation selection, but also the morphological destabilization of the GB that cannot be represented any longer by a straight line (or plane in 3D), but instead can also adopt highly convoluted shapes \cite{song2023cell}.
The sample thickness (1~mm) may also allow several grains to develop within the thickness, further contributing to the 3D growth competition.
This is prominently illustrated in Fig.~\ref{Fig:EBSD} by grain A growing in two separate locations of the sample.

In conclusion, even though the analysis of GB types allows a reasonable qualitative understanding of the discrepancy between simulation and experiments, we reckon that further work is required to obtain a fully conclusive comparison between simulation and experiments. 
Efforts in this direction are suggested in the conclusion section, most of which serve as the basis for ongoing work.

\section{Summary and perspectives}
\label{Sec:summary}

We presented an extensive numerical study of grain growth competition during directional solidification of hcp crystals, considering a Mg-1wt$\%$Gd alloy, and studied the effect of the temperature gradient over a wide range of $G$.
We also compared our results to classical and state-of-the-art theories for primary spacing selection and preferred growth direction.
We have shown that these theories, and the overall understanding of GB orientation selection in cubic (fcc,bcc) crystals may essentially be extended to the case of hcp crystals in the case of (nearly) two-dimensional configurations (i.e. within the basal plane), sometimes requiring scaling factors to account for the sixfold (rather than fourfold) symmetry.

Regarding grain growth competition, we have shown that, like in cubic crystals, converging grain boundaries and diverging grain boundaries with $\alpha_1\times\alpha_2>0$ (see Fig.~\ref{Fig:Imp}) tend to follow the orientation of the most favorably oriented grain, however with some amount of deviation due to relatively rare events of unusual overgrowth (at converging GBs) or side-branching (at diverging GBs) of favorably oriented dendrites.
The latter may actually be promoted by the fact that secondary branches, growing at a $60^\circ$ from the primary branch, actually face a higher temperature gradient than in fourfold (cubic) systems.
GB configurations exhibiting the highest stochastic variability are diverging GBs with $\alpha_1\times\alpha_2<0$, converging GBs close to a symmetric configuration with $\alpha_1=-\alpha_2$, and growth competition involving at least one grain with $\alpha_0\approx\pm30^\circ$, resulting in a dendrite-to-seaweed pattern transition.

Beyond these trends that could essentially be extrapolated from current knowledge on cubic crystals, we have also studied the effect of the temperature gradient $G$ on grain growth competition.
We found that, within the dendritic regime, a decrease of $G$ leads to the outcome of the growth competition approaching the theoretical geometrical limit (GL, Fig~\ref{Fig:FOG-GL_abs}b), whereas when increasing $G$ the selection map tends toward the favorably-oriented grain criterion (FOG, Fig~\ref{Fig:FOG-GL_abs}a), as suggested by previous studies on a subset of $(\alpha_1,\alpha_2)$ configuration in a cubic system \cite{dorari22}.
When increasing $G$ further (here for $G\geq2~$K/mm), the microstructure exhibits a transition from a dendritic to a cellular morphology. 
Then, as $G$ increases, the cell/dendrite growth direction tends to the temperature gradient direction ($\alpha\to0^\circ$) and, consequently, the selected GB orientation also significantly decreases to align close to that of $G$.

Finally, we also compared our PF prediction with experimental measurements on a directionally solidified thin samples of Mg-1wt$\%$Gd alloy. 
Within the few GB analyzed in details, we found a limited qualitative agreement, with notable discrepancies correlated with GB types expected to show the most stochastic variability according to our PF results.
Still, while we provide explanations for the discrepancy between experimental and computed GB angles (in particular three-dimensionality, stochasticity with small statistical sample size, and GB types), a fully conclusive and quantitative experimental comparison requires further research in the following directions.

Ongoing research includes gathering a broader statistic of GB orientation measurements, in thinner ($\approx 200~$\textmu m) samples, which will also be imaged using X-ray in situ radiography and 3D reconstructed tomography of the solidified samples.
Another important perspective could involve the use of fully 3D simulations. 
However, while studies of individually selected orientation configurations are achievable \cite{takaki18,song2023cell}, a full mapping of the 3D bicrystal orientation space, including multiple runs to account for GB stochasticity, seems out of reach even using the most high-performance PF implementations to date \cite{takaki18,elahi22}.
Emerging methods to explore the widely multidimensional space of alloys, processing parameters, and bi-crystal orientations, could include data-based tools, such as machine learning.
Regardless of whether further understanding will come from PF exploration similar to the present one or from data-based methods, a wider statistical dataset of experimental GB orientation under well-defined (e.g. $G$, $V$) conditions must be gathered. 
While this is not a conceptually challenging task, it could be tedious to obtain, and probably for this reason has not been undertaken, to our knowledge, by any group, for any crystal symmetry.

In summary, we have shown that several of the behaviors observed in grain growth competition of cubic crystals can be extrapolated to hcp crystals within their basal plane, provided the appropriate rescaling and re-evaluation of some fitting parameters.
We have also confirmed the transition from a GB selection from FOG to GL behavior with a decrease of temperature gradient, previously suggested for a limited set orientation configurations \cite{dorari22}, as well as the overall reduction of GB misorientation as $G$ increases beyond the onset of the dendritic-to-cellular morphological transition regime.
Experimental comparison, while encouraging, have highlighted the need for a broad statistically-relevant data set of GB orientation measurements from well-identified solidification conditions, as well as three-dimensional simulations, potentially supported by data-based approaches, in order to establish a more comprehensive understanding of grain growth competition during alloy solidification.

\section*{Acknowledgements}

This research was supported by the Spanish Ministry of Science within the scope of the HexaGB project (grant number RTI2018-098245) and the project MAD2D-CM of the Comunidad de Madrid, Recovery, Transformation and Resilience Plan, and NextGenerationEU from the European Union.
DT also gratefully acknowledges support from the Spanish Ministry of Science through a Ram\'on y Cajal
Fellowship (RYC2019-028233-I).

\appendix
\section{Anisotropy vector formalism for HCP symmetries}
\label{sec:app:anisovec}

In the present work, as in \cite{ghmadh14,boukellal18,boukellal21}, the evolution equation of the phase-field is written in terms of an auxiliary anisotropy vector $\vec{A}$, where 

\begin{equation}
\vec{\nabla}\cdot\vec{A} = \sum_{\alpha}\partial_\alpha\Big[|\vec{\nabla}\varphi|^2a_s\Big(\frac{\partial a_s}{\partial\varphi_\alpha}\Big)\Big],
\end{equation}

\noindent
where $\alpha\in\{x,y,z\}$, and the components of vector $\vec A=(A_x,A_y,A_z)$ directly depend on the anisotropy function of the surface energy, $a_s$, and can be computed for any arbitrary function $a_s(n_x,n_y,n_z)$. \\
Using the preconditioned phase-field 

\begin{equation}
\psi=\sqrt2 \tanh^{-1}({\varphi})
\label{psi}
\end{equation}

\noindent
and its first derivatives

\begin{equation}
\partial_\alpha\varphi=\frac{1}{\sqrt2}(1-\varphi^2)\partial_\alpha\psi,
\label{dpsi1}
\end{equation}          

\noindent
one obtains

\begin{equation}
A_\alpha= \frac{1-\varphi^2}{\sqrt{2}} |\vec{\nabla}\psi|^2a_s   \frac{\partial a_s}{\partial \psi_\alpha} .
\label{aalpha1}
\end{equation}

\noindent
Introducing the components 

\begin{equation}
n_\alpha= - \frac{\partial_\alpha \psi}{\vert\vec\nabla\psi\vert}
\end{equation} 

\noindent
of the normal vector $\vec{n}$ to the solid-liquid interface, the last term in Eq.~\eqref{aalpha1} becomes

\begin{equation}
\frac{\partial a_s}{\partial \psi_\alpha}=
\sum_\beta \frac{\partial a_s}{\partial n_\beta} \ \frac{\partial n_\beta}{\partial \psi_\alpha} = \frac{1}{|\vec\nabla \psi|}\sum_\beta (n_\alpha n_\beta - \delta_{\alpha,\beta}) \frac{\partial a_s}{\partial n_\beta}
\end{equation}

\noindent
with $\beta\in\{x,y,z\}$, and $\delta_{\alpha,\beta}=1$ if $\alpha=\beta$ or $0$ otherwise.
This leads to 

\begin{equation}
A_\alpha=\frac{ (1-\varphi^2)}{\sqrt2 } \vert \vec \nabla \psi \vert a_s
\sum_\beta (n_\alpha n_\beta-\delta_{\alpha,\beta})  \frac{\partial a_s}{\partial n_\beta}.
\label{Eq:Aalpha}
\end{equation}

\subsection*{\underline{Case of a HCP symmetry}}

\noindent 
For HCP crystals, the anisotropy function \cite{eiken10} is

\begin{align}
\gamma (\vec{n}) =~ \gamma _0 a_s(\vec{n}) =~& \gamma _0\Big[ 1+ \varepsilon_{20}y_{20}(\vec{n}) + \varepsilon_{40}y_{40}(\vec{n})  \nonumber\\
&+ \varepsilon_{60}y_{60}(\vec{n}) + \varepsilon_{66}y_{66}(\vec{n})          
\Big] ,
\label{eqAnis}
\end{align}  

\noindent where  

\begin{align}
y_{20} &=\sqrt{\frac{5}{16\pi}} (3n_z^2-1) \nonumber\\
&= a_{20}(3n_z^2-1); 
\\
y_{40} &= \frac{3}{16\sqrt{\pi}}(35n_z^4-30n_z^2+3) \nonumber\\
& = a_{40}(35n_z^4-30n_z^2+3);
\\
y_{60}&= \frac{\sqrt{13}}{32\sqrt{\pi}}(231nz^6-315n_z^4+105n_z^2-5)\nonumber \\
&= a_{60}(231n_z^6-315n_z^4+105n_z^2-5);\nonumber \\
\\
y_{66}&= \frac{\sqrt{6006}}{64\sqrt{\pi}}(n_x^6-15n_x^4n_y^2+15n_x^2n_y^4-n_y^6)\nonumber \\
&= a_{66}(n_x^6-15n_x^4n_y^2+15n_x^2n_y^4-n_y^6).\nonumber \\
\end{align}

\noindent 
Then 

\begin{align}
\frac{\partial a_s}{\partial n_x} &= 6\varepsilon_{66} a_{66}n_x (n_x^4 -10n_x^2n_y^2+5ny^4) ; 
\\
\frac{\partial a_s}{\partial n_y} &= -6\varepsilon_{66}a_{66}n_y(5n_x^4-10n_x^2n_y^2+n_y^4) ;
\\
\frac{\partial a_s}{\partial n_z}&= n_z\Big[6\varepsilon_{20}a_{20} + 20\varepsilon_{40}a_{40}(7n_z^2-3)\nonumber \\
&+ 6\varepsilon_{60}a_{60}(231n_z^4-210n_z^2+35)\Big].
\end{align}

\noindent This leads to the components of $\vec A$ 

\begin{align}
A_x =~& \frac{1-\varphi^2}{\sqrt{2}} | \vec{\nabla}\psi | a_s ~\times \nonumber \\
& \Big\{ 6\varepsilon_{66}a_{66}n_x(n_x^4-10n_x^2n_y^2+5n_y^4)(n_x^2-1)\nonumber \\
&-  6\varepsilon_{66}a_{66}n_y^2n_x(5n_x^4+10n_x^2n_y^2-n_y^4)\nonumber \\
&+ n_z^2n_x\Big[6\varepsilon_{20}a_{20} +20 \varepsilon_{40}a_{40}(7n_z^2-3) \nonumber \\
&+ 2\varepsilon_{60}a_{60}(693n_z^4-630n_z^2+105n_z)  \Big]
  \Big\} ;
\\
A_y =~& \frac{1-\varphi^2}{\sqrt{2}} | \vec{\nabla}\psi | a_s ~\times \nonumber \\
& \Big\{ 6\varepsilon_{66}a_{66}n_y \Big[n_x^2(n_x^4-10n_x^2n_y^2+5n_y^4)        \nonumber \\
&- (n_y^2-1)(5n_x^4-10n_x^2n_y^2+n_y^4) \Big]	\nonumber \\
&+	2n_z^2n_y \Big[3\varepsilon_{20}a_{20}+10\varepsilon_{40}a_{40}(7n_z^2-3)	 \nonumber \\
&+ \varepsilon_{60}a_{60}(693n_z^4-630n_z^2+105)  \Big]  \Big\} ;
\end{align}

\begin{align}
 \nonumber \\
 A_z  =~& \frac{1-\varphi^2}{\sqrt{2}} | \vec{\nabla}\psi | a_s ~\times \nonumber \\
& \Big\{ 6\varepsilon_{66}a_{66}n_z \Big[n_x^2(n_x^4-10n_x^2n_y^2+5n_y^4)        \nonumber \\
&- n_y^2(5n_x^4-10n_x^2n_y^2+n_y^4) \Big]	\nonumber \\
&+	2n_z (n_z^2-1)\Big[3\varepsilon_{20}a_{20}+10\varepsilon_{40}a_{40}(7n_z^2-3)	 \nonumber \\
&+ \varepsilon_{60}a_{60}(693n_z^4-630n_z^2+105)  \Big]  \Big\} .
\end{align}

\noindent
For convenience, these components can be reformulated as:

\begin{align}
A_x = ~&\frac{1-\varphi^2}{\sqrt{2}} | \vec{\nabla}\psi | a_s ~\times \nonumber \\
&  n_x\Big\{ -a_{66} \varepsilon_{66}n_y^2(30n_x^4-60n_x^2n_y^2+6n_y^4) \nonumber \\
&+ a_{66}\varepsilon_{66}(n_x^2-1)(6n_x^4-60n_x^2n_y^2+30n_y^4)  \nonumber \\
&+  n_z^2[6a_{20}\varepsilon_{20}+a_{40}\varepsilon_{40}(140n_z^2-60) \nonumber \\
&+ a_{60}\varepsilon_{60}(1386n_z^4-1260n_z^2+210)] 
  \Big\}
\\
A_y  = ~&\frac{1-\varphi^2}{\sqrt{2}} | \vec{\nabla}\psi | a_s ~\times \nonumber \\
& n_y\Big\{ a_{66} \varepsilon_{66}n_x^2(6n_x^4-60n_x^2n_y^2+30n_y^4) \nonumber \\
&- a_{66}\varepsilon_{66}(n_y^2-1)(30n_x^4-60n_x^2n_y^2+6n_y^4)  \nonumber \\
&+ n_z^2[6a_{20}\varepsilon_{20}+a_{40}\varepsilon_{40}(140n_z^2-60) \nonumber \\
&+ a_{60}\varepsilon_{60}(1386n_z^4-1260n_z^2+210)] 
  \Big\}
\\
A_z  =~& \frac{1-\varphi^2}{\sqrt{2}} | \vec{\nabla}\psi | a_s ~\times \nonumber \\
&  n_z\Big\{ a_{66} \varepsilon_{66}n_x^2(6n_x^4-60n_x^2n_y^2+30n_y^4) \nonumber \\
&-a_{66}\varepsilon_{66}n_y^2(30n_x^4-60n_x^2n_y^2+6n_y^4)  \nonumber \\
&+ (n_z^2-1)[6a_{20}\varepsilon_{20}+a_{40}\varepsilon_{40}(140n_z^2-60) \nonumber \\
&+a_{60}\varepsilon_{60}(1386n_z^4-1260n_z^2+210)] 
  \Big\}\end{align}

\noindent 
According to \cite{sun06},  

\begin{table}[h!]
\centering
\begin{tabular}{|c|c|c|c|c|}
\hline
$\varepsilon_{20}$&$\varepsilon_{40}$&$\varepsilon_{60}$&$\varepsilon_{66}$\\
\hline
$-0.026$&0&0&0.003\\
\hline
\end{tabular}
\end{table}

\noindent 
In two dimensions, the projection of Eq.~\eqref{eqAnis} on the xy-plane ($n_z = 0$) leads to

\begin{align}
\gamma (\vec{n}) = \gamma _0 a_s(\vec{n}) = \gamma _0\Big[ 1.008+\varepsilon_{66}y_{66}(\vec{n})          
\Big]
\label{eqAnisxy}
\end{align}  

\noindent 
and the components of $\vec A$, as they appear in the code, are

\begin{align}
A_x = ~&\frac{1-\varphi^2}{\sqrt{2}} | \vec{\nabla}\psi | a_s a_{66} \varepsilon_{66}  ~\times \nonumber \\
&  n_x\Big\{ -n_y^2(30n_x^4-60n_x^2n_y^2+6n_y^4) \nonumber \\
&+ (n_x^2-1)(6n_x^4-60n_x^2n_y^2+30n_y^4)  \Big\}
\\
A_y =~& \frac{1-\varphi^2}{\sqrt{2}} | \vec{\nabla}\psi | a_s a_{66}\varepsilon_{66}~\times \nonumber \\
& n_y\Big\{ n_x^2(6n_x^4-60n_x^2n_y^2+30n_y^4) \nonumber \\
&- (n_y^2-1)(30n_x^4-60n_x^2n_y^2+6n_y^4)  \Big\}
\end{align}  
  

\bibliographystyle{elsarticle-num}
\bibliography{biblio}

\end{document}